\documentclass[aps,prl,showpacs,twocolumn]{revtex4-2}
\usepackage{amsfonts}
\usepackage{amssymb}
\usepackage{amsmath}
\usepackage{graphicx}
\usepackage{epsfig}
\usepackage{subfigure}
\usepackage{appendix}
\usepackage{hyperref}

\hypersetup{hypertex=true,
	colorlinks=true,
	linkcolor=blue,
	urlcolor = blue,
	citecolor=blue}

\begin{document}

\title{Quantum Phase Transition in a Quantum Ising Chain at Nonzero
Temperatures}
\author{K. L. Zhang}
\author{Z. Song}
\email{songtc@nankai.edu.cn}
\affiliation{School of Physics, Nankai University, Tianjin 300071, China}

\begin{abstract}
We study the response of a thermal state of an Ising chain to a nonlocal
non-Hermitian perturbation, which coalesces the topological Kramer-like
degeneracy in the ferromagnetic phase. The dynamic responses for initial
thermal states in different quantum phases are distinct. The final state
always approaches its half component with a fixed parity in the
ferromagnetic phase but remains almost unchanged in the paramagnetic phase.
This indicates that the phase diagram at zero temperature is completely
preserved at finite temperatures. Numerical simulations for Loschmidt echoes
demonstrate such dynamical behaviors in finite-size systems. In addition, it
provides a clear manifestation of the bulk-boundary correspondence at
nonzero temperatures. This work presents an alternative approach to
understanding the quantum phase transitions of quantum spin systems at
nonzero temperatures.
\end{abstract}

\maketitle

\textit{Introduction}.---A conventional quantum phase transition (QPT) \cite%
{sachdev1999quantum} describes an abrupt change in matter at zero temperature. At
nonzero temperatures, the existence of quantum critical behavior depends on
the competition between thermal and quantum fluctuations. At higher
temperatures, thermal fluctuations conceal the quantum criticality, thus
leaving no residuals of quantum phase diagram at absolute zero temperature.
On the other hand, variations in a parameter across the critical point
induces a symmetry spontaneous breaking of the ground state. The underlying
mechanism is the degeneracy of the ground states. These features have been
demonstrated in a one-dimensional ($1$D) quantum Ising model with a
transverse field, which is exactly solvable, so as to be a unique paradigm
for understanding conventional QPTs. In recent works \cite{zhang2015topological, zhang2017majorana}%
, it turns out that the local order parameter and topological index can
coexist to characterize the QPT. 

In this Letter, we revisit the Ising model to investigate the existence of
QPT at nonzero temperatures---a seldom discussed topic. It is motivated from
the duality of the Kitaev model, which describes $1$-D spinless fermions
with superconducting $p$-wave pairing \cite{kitaev2001unpaired}. The Kitaev model is the
fermionized version of the familiar $1$-D transverse-field Ising model \cite%
{pfeuty1970one}, an easily solvable model exhibiting quantum criticality and QPT
with spontaneous symmetry breaking \cite{sachdev1999quantum}. Also, as the gene of
a Kitaev model, the Majorana lattice is the Su-Schrieffer-Heeger (SSH) model 
\cite{su1979solitons}, which has served as a paradigmatic example of a $1$-D system
supporting topological characteristic \cite{zak1989berry}. It manifests the key
features of topological order because the number of zero-energy levels and
edge states are immune to local perturbations \cite{asboth2016short}.

\begin{figure}[t]
\centering
\includegraphics[width=0.5\textwidth]{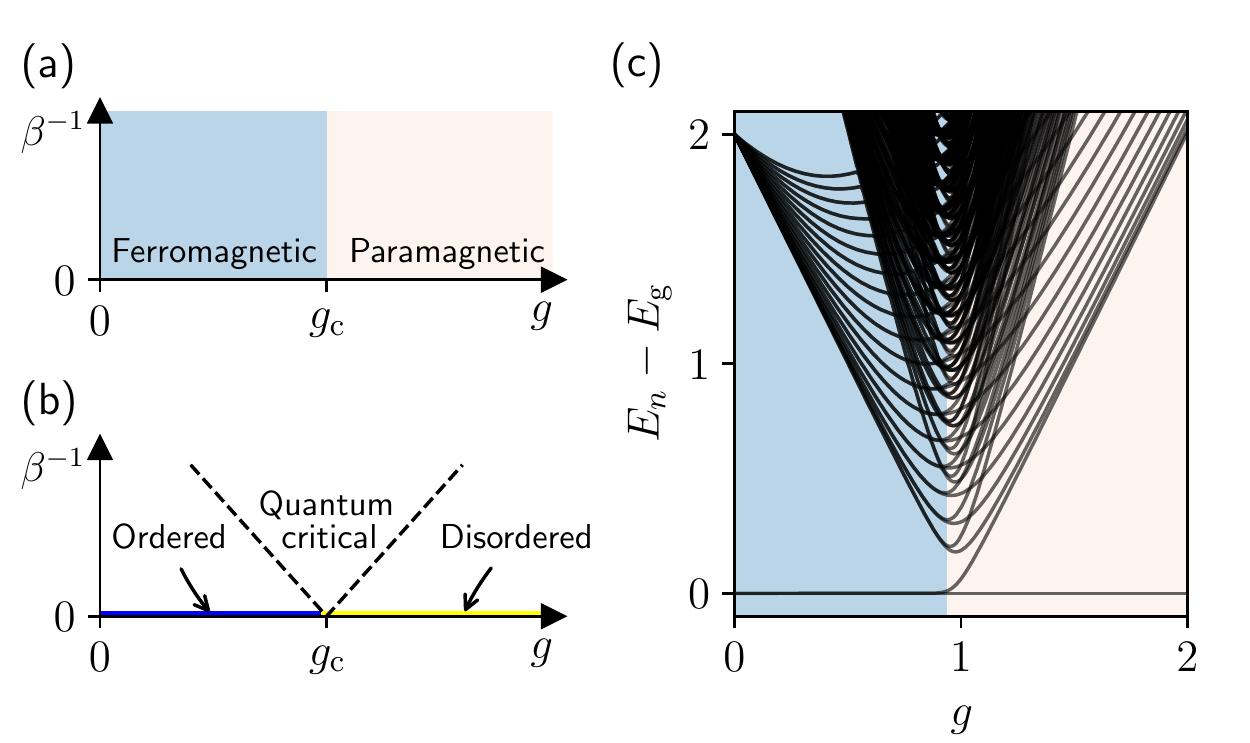}
\caption{(a) Phase diagram detected from the LEs in this work. 
(b) Phase diagram studied in term of correlation function in the work of Sachdev \textit{et al.}. 
Here $\beta^{-1}$ is the temperature and $g_{\mathrm{c}}$ is the quantum critical point. 
(c) Spectrum of the low-lying states for a finite quantum Ising chain
as a function of $g$, obtained numerically through exact diagonalization. $%
E_{\text{g}}$ is the ground-state energy. System parameters: $N=50$ and $J=1$%
. The energy gap closes at a quasicritical point, indicated by the boundary
of the two shaded areas. Notably, all energy levels become twofold
degeneracy simultaneously at one point, protected by the symmetry of the
quasi-zero-mode operator $D$.}
\label{fig1}
\end{figure}

At nonzero temperatures, there are various approaches to study quench dynamics 
of the Ising model and the XXZ model, for example, the form factor expansions \cite{dugave2013thermal, granet2020finite} and the quantum transfer matrix approaches \cite{suzuki1985transfer, andraschko2014dynamical}. A typical method 
for detecting QPT is to monitor the response of the ground state under a 
perturbation through the implementation of Loschmidt echo (LE)
and fidelity \cite{quan2006decay, zanardi2007mixed, cozzini2007quantum, heyl2013dynamical, abeling2016quantum, jafari2017loschmidt, mera2018dynamical}. 
Most perturbations applied to the Ising model are Hermitian terms, the
simplest example of which is the shift of the transverse field.
Nevertheless, since the discovery that a class of non-Hermitian Hamiltonians
could exhibit entirely real spectra \cite{mostafazadeh2002pseudo, bender2002complex, bender1998real, bender1999pt}, the non-Hermitian Hamiltonian is no longer a forbidden regime in
quantum mechanics. A certain type of non-Hermitian term may have exclusive
effects never before observed in a Hermitian system \cite{mostafazadeh2009spectral, longhi2014exceptional, jin2018incident, zhang2020dynamic}. More importantly, natural quantum systems such as cold
atom systems are intrinsically non-Hermitian because of spontaneous decay 
\cite{dalibard1992wave, dum1992monte, molmer1993monte, wiseman1996quantum, plenio1998quantum, lee2014heralded}. In this work, we
study the response of a thermal state of an Ising chain to a non-Hermitian
perturbation, which coalesces the topological Kramer-like degeneracy in the
ferromagnetic phase. We use LEs to measure the response and observe that
they are distinct for initial thermal states in different quantum phases.
The exceptional point (EP) drives a thermal state approaching to its half
component in the ferromagnetic phase but remain unchanged in the
paramagnetic phase. Numerical simulations for LE demonstrate such dynamical
behaviors in finite-size systems. In addition, it presents a clear
manifestation of the bulk-boundary correspondence at nonzero temperature.
The underlying mechanism is that within the ferromagnetic phase, the robust
degeneracy occurs not only in the ground states, but in all energy levels,
allowing the identification of the nature of quantum phases from a thermal
state. It indicates that the phase diagram at zero temperature is completely
preserved at finite temperatures [see Fig. \ref{fig1}(a)],
comparing to the phase diagram [see Fig. \ref{fig1}(b)] studied in terms of
correlation function in the work of Sachdev \textit{et al.} \cite%
{sachdev1999quantum, sachdev1997low}. This property promises the stable ground states,
and enables theoretical and experimental investigations of QPT through
dynamical control and testing. We present an alternative approach for
understanding the QPT of quantum spin systems at nonzero temperatures.

\textit{Model and degenerate spectrum}.---The model considered is the
transverse field Ising chain with open boundary condition, defined by the
Hamiltonian 
\begin{equation}
H=-J\sum_{j=1}^{N-1}\sigma _{j}^{x}\sigma _{j+1}^{x}+g\sum_{j=1}^{N}\sigma
_{j}^{z},  \label{H_Ising}
\end{equation}%
where $\sigma _{j}^{\alpha }$ ($\alpha =x,$ $y,$ $z$) are the Pauli
operators on site $j$ and parameter $g$ ($g\geqslant 0$) is the transverse
field strength. For simplicity, the following discussion assumes that $J=1$.
We first review some well-known model properties that are crucial to our
conclusion. The parity $p=\prod_{j=1}^{N}(-\sigma _{j}^{z})$ is determined
to be conservative; that is, $\left[ p,H\right] =0$ is always true.

The model with periodic boundary condition is exactly solvable and has been
well studied \cite{pfeuty1970one}. At zero temperature, QPT at $g=1$ separates a
ferromagnetic phase of the system ($g<1$) from a paramagnetic phase ($g>1$).
In general, model properties are not sensitive to the boundary condition in
thermodynamic limit. However, herein we consider the model with open
boundary condition, which notably possesses an exclusive symmetry in the
ferromagnetic phase $g<1$, and it is also the key point of this work. It can
be checked that in thermodynamic limit, we have a nonlocal operator \cite%
{Supplement} 
\begin{equation}
D=\frac{1}{2}\sqrt{1-g^{2}}\sum_{j=1}^{N}g^{j-1}D_{j},  \label{D}
\end{equation}%
with a position-dependent component%
\begin{equation}
D_{j}=\prod\limits_{l<j}\left( -\sigma _{l}^{z}\right) \sigma
_{j}^{x}-i\prod\limits_{l<N-j+1}\left( -\sigma _{l}^{z}\right) \sigma
_{N-j+1}^{y},  \label{Dj}
\end{equation}%
(where $i=\sqrt{-1}$), satisfying the commutation relations 
\begin{equation}
\lbrack D,H]=[D^{\dag },H]=0,  \label{symmetry}
\end{equation}%
that can be regarded as a symmetry of the system. In addition, the relations 
$\{D,D^{\dag }\}=1$ and $D^{2}=(D^{\dag })^{2}=0$ \cite{Supplement} suggest
that $D$ is a fermion operator, which can be related to the edge operator of
the Kitaev chain \cite{kitaev2001unpaired} $D\rightarrow \frac{1}{2}\sqrt{1-g^{2}}\sum_{j=1}^{N}[%
( g^{j-1}+g^{N-j}) c_{j}^{\dagger }+( g^{j-1}-g^{N-j})c_{j}]$ 
(where $c_{j}$ is a fermion operator) by the Jordan-Wigner
transformation \cite{jordan1993paulische}. Importantly, such a symmetry is a little special, because it
is contingent on the following conditions: $g<1$, a large $N$ limit, and
open boundary. Particularly, operator $D$ is nonuniversal and
Hamiltonian dependent because it contains the parameter $g$ from the
Hamiltonian. The first two conditions accord with the symmetry breaking
mechanism of QPT \cite{sachdev1999quantum}. Actually, the commutation relations in
Eq. (\ref{symmetry}) guarantee the existence of eigenstate degeneracy.
Specifically, there is a set of degenerate eigenstates $\left\{ \left\vert
\psi _{n}^{+}\right\rangle ,\left\vert \psi _{n}^{-}\right\rangle \right\} $
of $H$ with eigenenergy $E_{n}$, in two invariant subspaces, i.e., $%
H\left\vert \psi _{n}^{\pm }\right\rangle =E_{n}\left\vert \psi _{n}^{\pm
}\right\rangle $ and $p\left\vert \psi _{n}^{\pm }\right\rangle =\pm
\left\vert \psi _{n}^{\pm }\right\rangle $. Figure \ref{fig1}(c) presents the
spectrum of the low-lying states, which possess distinct degenerate
structures in two phases. Furthermore, we have the relations%
\begin{equation}
D\left\vert \psi _{n}^{+}\right\rangle =\left\vert \psi
_{n}^{-}\right\rangle ,D^{\dag }\left\vert \psi _{n}^{-}\right\rangle
=\left\vert \psi _{n}^{+}\right\rangle ,D^{\dag }\left\vert \psi
_{n}^{+}\right\rangle =D\left\vert \psi _{n}^{-}\right\rangle =0,
\label{mapping}
\end{equation}%
in the ferromagnetic phase. We refer to this property as topological
Kramers-like degeneracy for two reasons: (i) the twofold degeneracy lies in
the full spectrum, and (ii) it is invariant in the presence of random,
position-dependent deviation on the field $g$, where a new operator $D$ is
redefined accordingly \cite{Supplement}. Because of this property, operator $%
D$ plays an important role in the quench dynamics, as demonstrated in the
following section.

\textit{Non-Hermitian perturbation and EP dynamics}.---In general, a
Hermitian perturbation can lift the degeneracy. However, a non-Hermitian
perturbation may take a surprising effect. A fascinating phenomenon is the
coalescence of two degenerate states, which supports exclusive dynamics
never occurs in a Hermitian system. Such degeneracy-related dynamics
differentiates the quantum phases at any temperature, not only in the ground
states. To this end, we introduce operator $D$ into the post-quench
Hamiltonian $\mathcal{H}$ by treating it as a perturbation 
\begin{equation}
\mathcal{H}=H+\kappa D,  \label{H1}
\end{equation}%
with $\kappa \ll g$. For a system in the ferromagnetic phase, where $0<g<1$,
any pair of degenerate eigenstates $\left( \left\vert \psi
_{n}^{+}\right\rangle ,\left\vert \psi _{n}^{-}\right\rangle \right)$ with
energy $E_{n}$ spans a diagonal block with the sub-Hamiltonian 
\begin{equation}
\mathcal{H}_{n}=\left( 
\begin{array}{cc}
E_{n} & 0 \\ 
\kappa & E_{n}%
\end{array}%
\right) ,
\end{equation}%
which has a Jordan block structure. This means that in the ferromagnetic
phase, the degenerate spectrum becomes an exceptional spectrum with a set of
coalescing states $\left\{ \left\vert \psi _{n}^{\mathrm{c}}\right\rangle
\right\} =\left\{ \left\vert \psi _{n}^{-}\right\rangle \right\} $ when the
non-Hermitian term $\kappa D$ is introduced. The diagonal Jordan block is
exact for any values of $\kappa $. By contrast, for a system in the
paramagnetic phase, $\kappa D$ does not considerably affect the energy
levels of $H$ when $g$ is much larger than $1$ ($H\approx
g\sum_{j=1}^{N}\sigma _{j}^{z}$ in this case); this is because the gap
between energy levels with different parities is at least in the order of $g$
[see Fig. \ref{fig1}(c)].

On the basis of this analysis, the dynamics in the ferromagnetic phase is
governed by the time evolution operator%
\begin{equation}
U(t)=\exp (-i\mathcal{H}t)=\prod_{n}U_{n}(t),
\end{equation}%
where the time evolution operator in the $n$th subspace has the form $%
U_{n}(t)$ $=\exp (-i\mathcal{H}_{n}t)$ $=\exp (-iE_{n}t)$ $\left[ 1-i(%
\mathcal{H}_{n}-E_{n})t\right] $ based on the identity $\left( \mathcal{H}%
_{n}-E_{n}\right) ^{2}=0$ for $g<1$. The dynamics of a pure initial state
are then clarified, as given in $U_{n}(t)\left( a\left\vert \psi
_{n}^{-}\right\rangle +b\left\vert \psi _{n}^{+}\right\rangle \right) $ $%
=\exp (-iE_{n}t)\left[ \left( a-itb\kappa \right) \left\vert \psi
_{n}^{-}\right\rangle +b\left\vert \psi _{n}^{+}\right\rangle \right] $. The
action of $U_{n}(t)$ over a long period projects any pure initial state on
the component $\left\vert \psi _{n}^{-}\right\rangle $, which is completely
different from that in the paramagnetic phase. These features allow us to
observe significantly different dynamical behaviors for a initial thermal
state.

\textit{QPT at nonzero temperatures}.---We have observed that the difference
between spectra in two regions not only lies in the ground states but also
the full spectrum. This results in the exclusive EP dynamics for the initial
state involving any excited eigenstates in the ferromagnetic phase, from
which two phases at finite temperatures can be identified. Notably,
bulk-boundary correspondence can manifest at nonzero temperatures. In the
following, we focus on the dynamics of a initial thermal state with density
matrix $\rho \left( 0\right) =e^{-\beta H}/\mathrm{Tr}e^{-\beta H}$ at
temperature $\beta $ for a system (pre-quench Hamiltonian) $H$ under a
quenched non-Hermitian Hamiltonian $\mathcal{H}=H+\kappa H^{\prime }$ where $%
H^{\prime } $ is non-Hermitian, and $\kappa $ is real.

As mentioned, operator $D$ is $g$ dependent, and a matching $D$ in the
perturbation leads to an exact EP. Nevertheless, operator $D_{j}$ (or $%
D_{j}^{\dag }$) still takes the role to switch the parity of an eigenstate
and forms a Jordan block approximately for a sufficiently small $\kappa $. 
Operators $D$ and $D_{j}$ ($j\in \lbrack 1,N]$) are nonlocal combinations
of spin operators $\left\{ \sigma _{j}^{x}\right\} $ and $\left\{ \sigma_{j}^{y}\right\} $ 
for a quantum spin system, and $D_{1}$ is the main component of $D$. 
We consider two cases of $H^{\prime }$ where it is (i) a dominant term of
operator $D$ (i.e., $H^{\prime }=D_{1}$) and (ii) position dependent (i.e., $%
H^{\prime }=D_{j}$). After the quench, the time evolution of the thermal
state obeys the equation 
\begin{equation}
i\frac{\partial }{\partial t}\rho \left( t\right) =\mathcal{H}\rho \left(
t\right) -\rho \left( t\right) \mathcal{H}^{\dag },
\end{equation}%
which admits the formal solution 
\begin{equation}
\rho \left( t\right) =e^{-i\mathcal{H}t}\rho \left( 0\right) e^{i\mathcal{H}%
^{\dag }t}.
\end{equation}%
Unlike the Hermitian case, the time evolution of the density matrix is no
longer unitary. Thus, in the following numerical calculation, we normalize $%
\rho \left( t\right) $ by taking \cite{brody2012mixed, kawabata2017information} 
\begin{equation}
\rho \left( t\right) =e^{-i\mathcal{H}t}\rho \left( 0\right) e^{i\mathcal{H}%
^{\dag }t}/\mathrm{Tr}\left[ e^{-i\mathcal{H}t}\rho \left( 0\right) e^{i%
\mathcal{H}^{\dag }t}\right] .
\end{equation}

To characterize the degree of distinguishability between the initial state $%
\rho \left( 0\right) $ and evolved state $\rho \left( t\right) $, we
introduce the LE 
\begin{equation}
L\left( t\right) =\left[ \mathrm{Tr}\sqrt{\sqrt{\rho \left( 0\right) }\rho
\left( t\right) \sqrt{\rho \left( 0\right) }}\right] ^{2},  \label{Lt}
\end{equation}%
also known as the Uhlmann fidelity \cite{uhlmann1976transition, jacobson2011unitary}. The value of 
$L\left( t\right) $ after a sufficient period can be estimated intuitively.
In general, an initial mixed state $\rho \left( 0\right) $ contains 
components of two parities. In the ferromagnetic phase, the
component with a certain parity of the thermal state $\rho \left( t\right) $
is dominant because of EP dynamics, and in large $t$ limits, the LE $L\left(
t\right) $ approaches $0.5$ . In the paramagnetic phase, a non-Hermitian
perturbation does not substantially affect the dynamics; this is expressed
by $L\left( t\right) \approx L\left( 0\right) =1$. We now numerically
demonstrate the decay behavior of $L\left( t\right) $ within a short period.

\begin{figure}[t]
	\centering
	\includegraphics[width=0.5\textwidth]{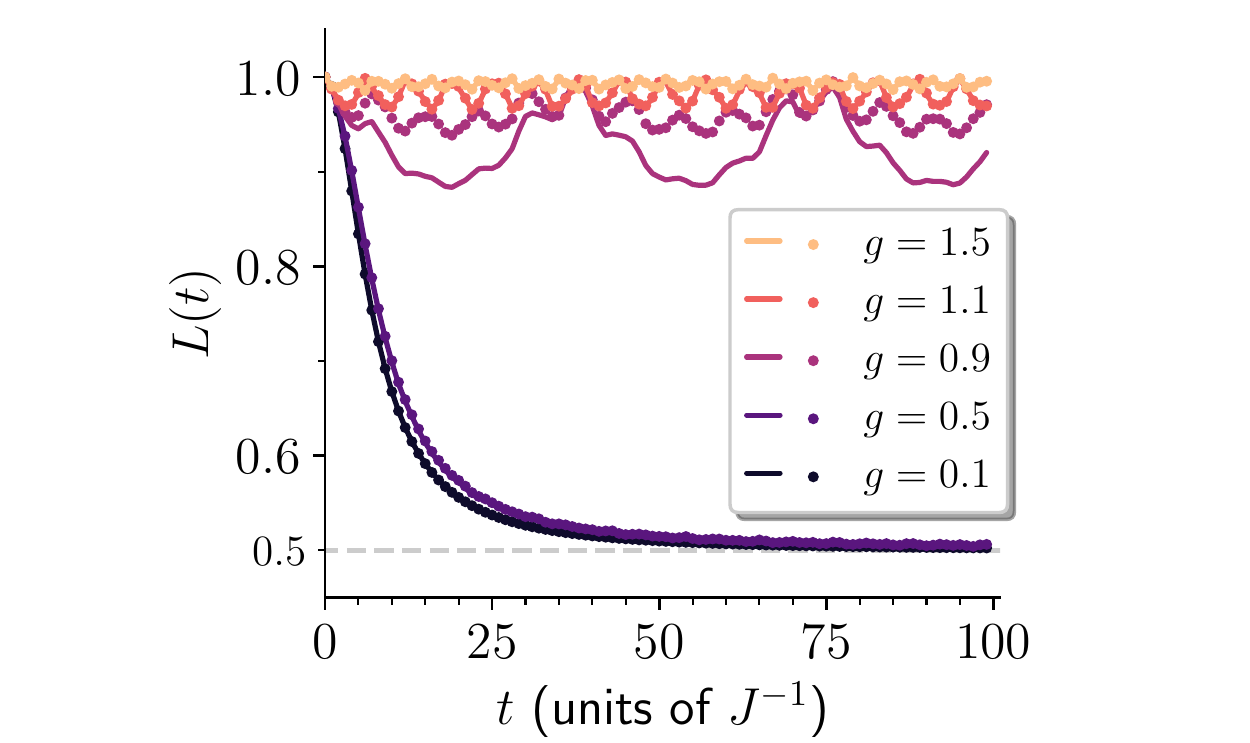}
	\caption{LEs of different $g$ values. The lines and dots represent the LEs
		for $\protect\beta =5$ and $\protect\beta =10$, respectively. Other
		parameters: $N=10$, $\protect\kappa =0.1$, and $J=1$. The profiles of the LEs
		in the two regions are distinct, independent of the temperature of the
		initial thermal states, and converge to $1.0$ and $0.5$, respectively. }
	\label{fig2}
\end{figure}

First, we consider the quench dynamics under the postquench Hamiltonian $%
\mathcal{H}=H+\kappa D_{1}$. We conduct numerical simulations for $L\left(
t\right) $ for the initial state $\rho \left( 0\right) $ at different phases
in the finite system. The computations are performed using a uniform mesh in
the time discretization for the Hamiltonian $\mathcal{H}$. As mentioned, the
spectral degeneracy is dependent on a large $N$ limit. However,
a sufficiently small $g$ still leads to perfect quasidegeneracy in
finite-size systems \cite{Supplement}. Consistent with our prediction, the
numerical results of LEs in Fig. \ref{fig2} are insensitive to temperature
and tend towards different values in different phases.

\begin{figure}[t]
\centering
\includegraphics[width=0.5\textwidth]{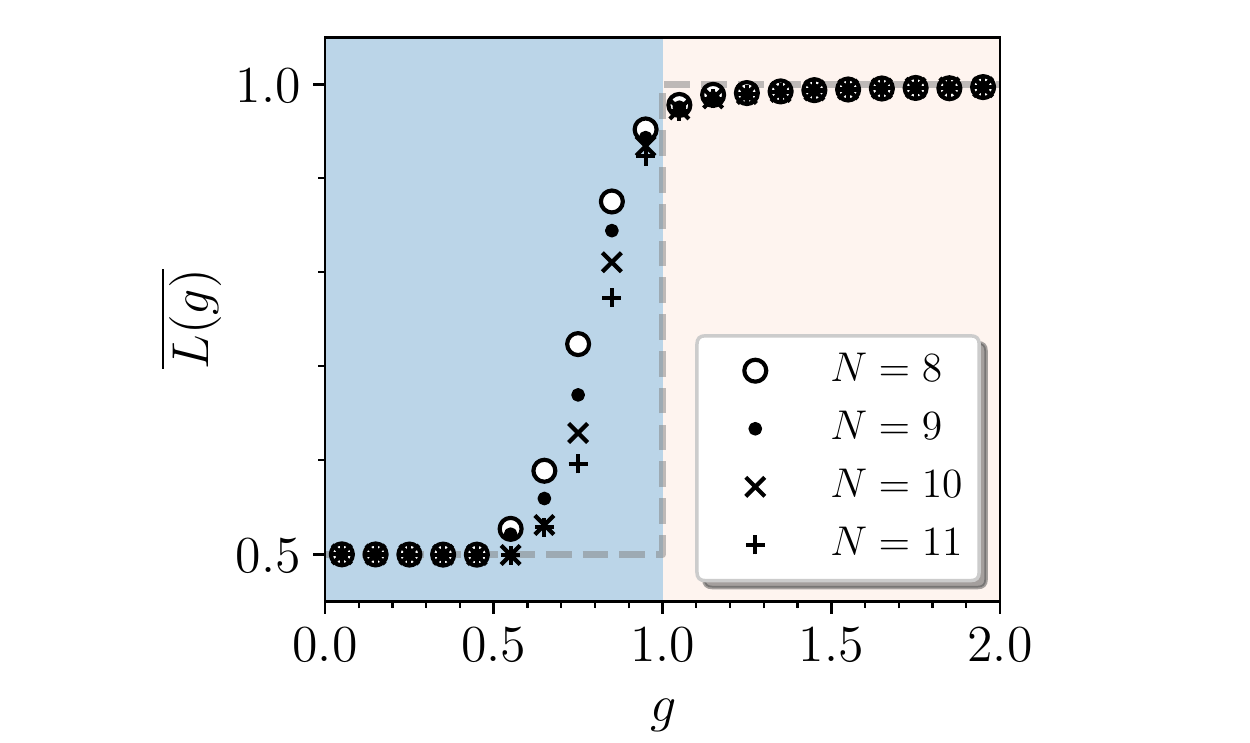}
\caption{Average LEs as functions of $g$ when $N=8$, $9$, $10$, and $11$. The
dashed line represents the ideal average LEs expected for large $N$ limits.
Here we set $\protect\tau =500$ and $T=500$. Other parameters: $\protect%
\kappa =0.1$, $J=1$, and $\protect\beta =1$. It indicates that as $N$
increases, the plots have the trends to the prediction in the thermodynamic
limit.}
\label{fig3}
\end{figure}

To determine the effect of $g$, we introduce an average LE in the time
interval $[\tau ,\tau +T]$, defined as follows: 
\begin{equation}
\overline{L}=\frac{1}{T}\int_{\tau }^{\tau +T}L\left( t\right) dt,
\end{equation}%
where $\tau \gg 1$. Average LEs as functions of parameter $g$ for different $%
N$ values are plotted in Fig. \ref{fig3}. When $N$ is larger, the average LE
is closer to the ideal values that are expected in the thermodynamic limit. This
indicates that the LEs can be used to identify the quantum phase diagram at
nonzero temperatures even in small size systems.

\begin{figure}[t]
\centering
\includegraphics[width=0.5\textwidth]{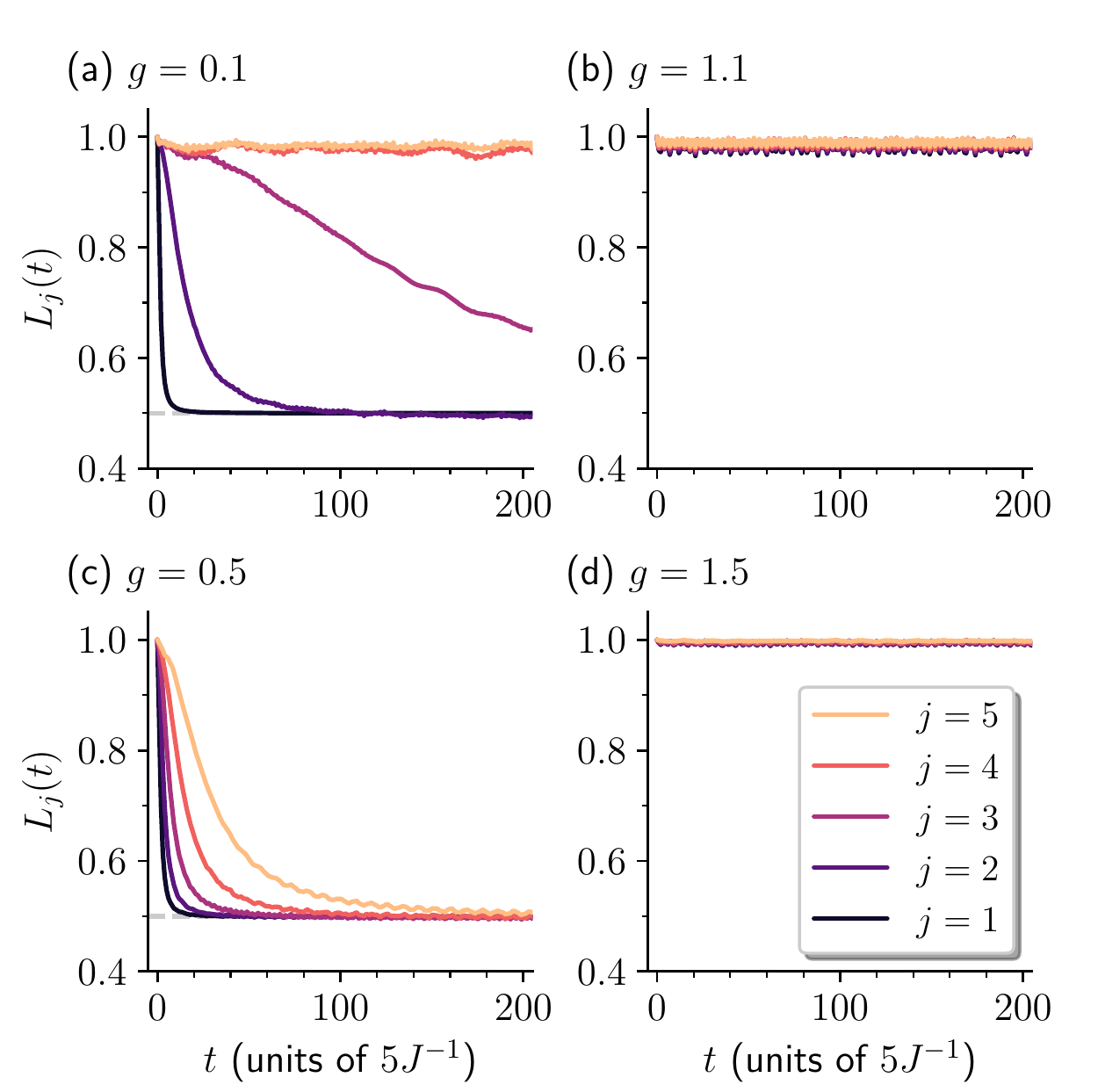}
\caption{Simulation results for LEs under the postquench Hamiltonian (%
\protect\ref{H2}) for different $j$ values. (a) and (c) LEs in the
ferromagnetic phase when $g=0.1$ and $0.5$. (b) and (d) LEs in the
paramagnetic phase when $g=1.1$ and $1.5$. Other parameters: $N=10$, $J=1$, $%
\protect\kappa =0.1$, and $\protect\beta =1$. The LE decays rapidly to $0.5$
at the end of the chain in the ferromagnetic phase, whereas it remains at
one in the paramagnetic phase. In the case (a), where $g$ is small, the LEs
in the middle do not decay but remain near $1.0$, which implies the LE
behavior in long chains. This is a clear manifestation of the bulk-boundary
correspondence at nonzero temperatures.}
\label{fig4}
\end{figure}

Second, we investigate the bulk-boundary correspondence at nonzero
temperatures through quench dynamics. Consider the post-quench Hamiltonian
with the form 
\begin{equation}
\mathcal{H}=H+\kappa D_{j},  \label{H2}
\end{equation}%
where $D_{j}$, defined in Eq. (\ref{Dj}), is the component of operator $D$.
In this case, the LE is denoted by $L_{j}\left( t\right) $. The long-term
behavior of $L_{j}\left( t\right) $ when $j>1$ is expected to be similar to $%
L\left( t\right) $ of the postquench Hamiltonian in the first case. We are
interested in the dependence of $L_{j}\left( t\right) $ in different phases
on position over a short period. The numerical simulation results are
plotted in Fig. \ref{fig4}. We can see that, (i) in the case of $g<1$, $%
L_{j}\left( t\right) $ tends towards $0.5$ for the $j$ near the end and
decays more rapidly as $j$ approaches the boundary. By contrast, in the case
of $g>1$, $L_{j}\left( t\right) $ remains at $1.0$ for all $j$. And (ii) in
the case of $g=0.1$, the LEs in the middle do not decay but remain near a
value of one. The expression of $D$ indicates that in the ferromagnetic
phase $g<1$, small $g$ values enhances the edge effect compared with the
case of large $g$ \cite{SupplementC}. This suggests that similar effects can be observed in any
case of a sufficiently long chain where $g<1$. In addition, this
demonstrates that $L_{j}\left( t\right) $ defines the bulk-boundary
correspondence for Ising chains at nonzero temperatures.

\textit{Discussion}.--- In summary, we extended the quantum phase diagram
for Ising chains from zero to nonzero temperatures. The degeneracy spectrum
of the system in the ferromagnetic phase, which arises from Majorana zero
modes, is the crux of our conclusion. Such nonzero-temperature QPT can be
detected through the response of a thermal state to a nonlocal non-Hermitian
perturbation on the Ising chain. The non-Hermiticity of the perturbation
dynamically amplifies the difference between two quantum phases. The
EP dynamics for coalescing states have no counterpart in the Hermitian
regime and allow distinct responses for initial thermal states in the two
quantum phases. Numerical simulations for LE also provides a clear
manifestation of the bulk-boundary correspondence at nonzero temperatures in
the quantum spin system. This is an alternative approach for understanding
the QPT of quantum spin systems at nonzero temperatures.
The possible experimental implementations to verify our results would be a 
diamond system \cite{huang2011observation} and polycrystalline adamantane system \cite{sanchez2020perturbation}, 
where the dynamical behaviors in quantum spin systems at nonzero temperature were observed.

Several points should be addressed before ending this Letter. (i) The
nonlocal factor $\prod\nolimits_{l<j}\left( -\sigma _{l}^{z}\right) $ in
operator $D_{j}$ has a crucial role in the simulations; when it is omitted,
the EP cannot appear again. (ii) In the presence of disordered parameters $J$
and $g$ in the Hamiltonian, the observed results still hold \cite{Supplement}%
. (iii) The approach based on thermal state fidelity can also be applied to
the non-Hermitian model $\mathcal{H}(g)=H+\kappa D$. Any thermal state
always has a fixed parity in the ferromagnetic phase, while it has half
component with each party in the paramagnetic phase. This leads to a sudden
drop in the thermal fidelity at the critical point.

\acknowledgments This work was supported by the National Natural Science Foundation of China
(under Grant No. 11874225).

\clearpage
\begin{widetext}
\section{Supplemental Material}

\begin{center}
	K. L. Zhang and Z. Song*\\[2pt]
	\textit{School of Physics, Nankai University, Tianjin 300071, China}\\
	*songtc@nankai.edu.cn
\end{center}

\setcounter{equation}{0} \renewcommand{\theequation}{S\arabic{equation}} %
\setcounter{figure}{0} \renewcommand{\thefigure}{S\arabic{figure}} %
\setcounter{secnumdepth}{3}

In this Supplemental Material, we present \ref{A}. Derivation of 
the operator $D$: uniform case; \ref{B}. Derivation of the operator $D$: 
disordered case; and \ref{C}. Approximate calculation of the Loschmidt echo in a larger $N$.

\subsection{Derivation of the operator $D$: uniform case}
\label{A} 
Starting from the Ising chain Hamiltonian $H$ with $J=1$ in the
Letter, one can perform the Jordan-Wigner transformation \cite{jordan1993paulische} 
\begin{eqnarray}
	\sigma _{j}^{x} &=&\prod\limits_{l<j}\left( 1-2c_{l}^{\dagger }c_{l}\right)
	\left( c_{j}+c_{j}^{\dagger }\right) ,  \notag \\
	\sigma _{j}^{z} &=&2c_{j}^{\dag }c_{j}-1,
\end{eqnarray}%
to replace the Pauli operators by the fermionic operators $c_{j}$. The
Hamiltonian is transformed to the Kitaev model \cite{kitaev2001unpaired}%
\begin{equation}
	H_{\text{Kitaev}}=-\sum_{j=1}^{N-1}\left( c_{j}^{\dagger
	}c_{j+1}+c_{j}^{\dag }c_{j+1}^{\dag }\right) +\text{\textrm{H.c.}}%
	+g\sum_{j=1}^{N}\left( 2c_{j}^{\dagger }c_{j}-1\right) .
\end{equation}%
To get the solution of the model, we introduce the Majorana fermion
operators $a_{j}=c_{j}^{\dagger }+c_{j},b_{j}=-i\left( c_{j}^{\dagger
}-c_{j}\right) ,$which satisfy the commutation relations $\left\{
a_{j},a_{j^{\prime }}\right\} =2\delta _{j,j^{\prime }},\left\{
b_{j},b_{j^{\prime }}\right\} =2\delta _{j,j^{\prime }},\left\{
a_{j},b_{j^{\prime }}\right\} =0.$ Then the Majorana representation of the
original Hamiltonian is%
\begin{equation}
	H_{\text{M}}=-\frac{i}{2}\sum_{j=1}^{N-1}b_{j}a_{j+1}-\frac{i}{2}%
	g\sum_{j=1}^{N}a_{j}b_{j}+\text{\textrm{H.c.,}}  \label{H_SSH}
\end{equation}%
the core matrix of which is that of a $2N$-site Su-Schrieffer-Heeger (SSH)
chain in single-particle invariant subspace. Based on the exact
diagonalization result of the SSH chain, the Hamiltonian $H_{\text{Kitaev}}$%
\ can be written as the diagonal form 
\begin{equation}
	H_{\text{Kitaev}}=\sum_{n=1}^{N}\varepsilon _{n}\left( d_{n}^{\dagger }d_{n}-%
	\frac{1}{2}\right) .
\end{equation}%
Here $d_{n}$\ is a fermionic operator, satisfying $\{d_{n},d_{n^{\prime
}}\}=0,$ and $\{d_{n},d_{n^{\prime }}^{\dag }\}=\delta _{n,n^{\prime }}$. On
the other hand, we have the relations%
\begin{equation}
	\left[ d_{n},H_{\text{Kitaev}}\right] =\varepsilon _{n}d_{n},\left[
	d_{n}^{\dagger },H_{\text{Kitaev}}\right] =-\varepsilon _{n}d_{n}^{\dagger },
	\label{dHcommu}
\end{equation}%
which result in the mapping between the eigenstates of $H_{\text{Kitaev}}$.
Direct derivation show that, for an arbitrary eigenstate $\left\vert \psi
\right\rangle $ of $H_{\text{Kitaev}}$ with eigenenergy $E$, i.e., $H_{\text{%
		Kitaev}}\left\vert \psi \right\rangle =E\left\vert \psi \right\rangle $,
state $d_{n}\left\vert \psi \right\rangle $\ $\left( d_{n}^{\dag }\left\vert
\psi \right\rangle \right) $\ is also an eigenstate of $H_{\text{Kitaev}}$\
with the eigenenergy $E-\varepsilon _{n}$ $\left( E+\varepsilon _{n}\right) $%
, i.e.,%
\begin{equation}
	H_{\text{Kitaev}}\left( d_{n}\left\vert \psi \right\rangle \right) =\left(
	E-\varepsilon _{n}\right) \left( d_{n}\left\vert \psi \right\rangle \right)
\end{equation}%
and%
\begin{equation}
	H_{\text{Kitaev}}\left( d_{n}^{\dag }\left\vert \psi \right\rangle \right)
	=\left( E+\varepsilon _{n}\right) \left( d_{n}^{\dag }\left\vert \psi
	\right\rangle \right) ,
\end{equation}%
if $d_{n}\left\vert \psi \right\rangle \neq 0$ $\left( d_{n}^{\dag
}\left\vert \psi \right\rangle \neq 0\right) $.

In large $N$ limit, and within the topologically nontrivial region $%
\left\vert g\right\vert <1$ ($g\neq 0$), the edge modes appear with $%
\varepsilon _{N}=0$\ and the edge operator $d_{N}$\ can be expressed as%
\begin{equation}
	d_{N}=\frac{1}{2}\sqrt{1-g^{2}}\sum_{j=1}^{N}\left[ \left(
	g^{j-1}+g^{N-j}\right) c_{j}^{\dagger }+\left( g^{j-1}-g^{N-j}\right) c_{j}%
	\right] ,
\end{equation}%
i.e., $d_{N}$\ is a linear combination of particle and hole operators of
spinless fermions $c_{j}$\ on the edge, and we have $\left[ d_{N},H_{\text{%
		Kitaev}}\right] =\varepsilon _{N}d_{N}=0$. Furthermore, applying the inverse
Jordan-Wigner transformation, $d_{N}$ can be expressed as the combination of
spin operators, 
\begin{eqnarray}
	D &=&\frac{1}{2}\sqrt{1-g^{2}}\sum_{j=1}^{N}\prod\limits_{l<j}\left( -\sigma
	_{l}^{z}\right) \left( g^{j-1}\sigma _{j}^{x}-ig^{N-j}\sigma _{j}^{y}\right)
	\notag \\
	&=&\frac{1}{2}\sqrt{1-g^{2}}\sum_{j=1}^{N}g^{j-1}D_{j},
\end{eqnarray}%
where $D_{j}=\prod_{l<j}\left( -\sigma _{l}^{z}\right) \sigma
_{j}^{x}-i\prod_{l<N-j+1}\left( -\sigma _{l}^{z}\right) \sigma _{N-j+1}^{y}$.

In fact, $d_{N}$\ and $D$\ are identical, but only in different
representations. Thus, from $\left[ d_{N},H_{\text{Kitaev}}\right] =0$, we
have 
\begin{equation}
	\left[ D,H\right] =\left[ D^{\dag },H\right] =0,  \label{commu1}
\end{equation}%
which lead to the degeneracy of the eigenstates. Here we would
like to point out that the spectral degeneracy is dependent on a large $N$
limit. Nevertheless, a sufficiently small $g$ still leads to perfect
quasidegeneracy in finite-size systems, since from the exact diagonalization
result of a finite-size SSH chain we have $\varepsilon _{N}\thicksim g^{N}$.
Furthermore, from the canonical commutation relations $\left\{
d_{N},d_{N}^{\dag }\right\} =1$ and $\left\{ d_{N},d_{N}\right\} =0$, we
have 
\begin{equation}
	\left\{ D,D^{\dag }\right\} =1,D^{2}=\left( D^{\dag }\right) ^{2}=0.
	\label{commu2}
\end{equation}

Operators $D$\ and $D_{j}$\ ($j\in \lbrack 1,N]$) are nonlocal combinations
of spin operators $\left\{ \sigma _{j}^{x}\right\} $\ and $\left\{ \sigma
_{j}^{y}\right\} $\ for a quantum spin system, and $D_{1}$ is the main component 
of $D$. Operator $D=\frac{1}{2}\sqrt{1-g^{2}}\sum_{j}g^{j-1}D_{j}$ commutes with the Hamiltonian and acts as a
raising (or lowering) operator for two degeneracy eigenstates for $g<1$.
Meanwhile, they are essentially spinless fermion operators for the fermion
representation of the quantum spin system.

The mechanism of the nonlocal non-Hermitian perturbation in the Letter 
is based on an exclusive feature of a non-Hermitian system,
which is the existence of exceptional point (EP). Unlike the degeneracy in a
Hermitian system, two or more eigenstates coalesce into a single eigenstate.
Notably, it supports a special dynamics, which has no counterpart in the
Hermitian regime.
Such an approach can be applied to other models, which possess degenerate
spectrum. In general, such a degeneracy is originated from a symmetry, or a
fermionic operator commuting with the Hamiltonian. If such an operator is
non-Hermitian, then the Jordan block is formed, which allows the EP 
dynamics to demonstrate the existence of the degenerate spectrum.
Technically speaking, this operator can be solved in the fermionic
representation, as the edge operator of the fermionic chain.

\subsection{Derivation of the operator $D$: disordered case}
\label{B}
For the Ising chain with position-dependent random $J_{j}$ and $g_{j}$,
i.e., $H=-\sum_{j=1}^{N-1}J_{j}\sigma _{j}^{x}\sigma
_{j+1}^{x}+\sum_{j=1}^{N}g_{j}\sigma _{j}^{z}$, the operator $D$ still
exists. In this case, one can perform the above procedure and solve the Schr%
\"{o}dinger equation for the corresponding SSH chain with random hopping in
single-particle invariant subspace \cite{asboth2016short}. We have the following
solution:%
\begin{equation}
	D=\frac{1}{2}\sum_{j=1}^{N}\prod\limits_{l<j}\left( -\sigma _{l}^{z}\right)
	\left( h_{j}^{+}\sigma _{j}^{x}-ih_{j}^{-}\sigma _{j}^{y}\right) ,
\end{equation}%
where%
\begin{eqnarray}
	h_{j}^{+} &=&h_{1}^{+}\prod\limits_{m=1}^{j-1}\frac{g_{m}}{J_{m}},  \notag \\
	h_{j}^{-} &=&h_{N}^{-}\frac{g_{N}}{J_{j}}\prod\limits_{m=j+1}^{N-1}\frac{%
		g_{m}}{J_{m}},
\end{eqnarray}%
and $h_{1}^{+}$ ($h_{N}^{-}$) is determined by the normalization condition $%
\sum_{j=1}^{N}\left\vert h_{j}^{\pm }\right\vert ^{2}=1.$ The solution of $D$
is robust against disordered perturbation and the corresponding energies $\varepsilon _{N}$ of 
the edge modes are still exponentially small in $N$ under the condition of
the average value of $J_{m}$ is stronger than the average value of $g_{m}$ 
\cite{asboth2016short}. Then it can be checked that the commutation relations in Eqs.
(\ref{commu1}) and (\ref{commu2}) still hold for the operator $D$ with
disordered perturbation in large $N$ limit.
This leads to the robust
degeneracy of the eigenstates, suggesting that the observed results in the Letter 
still hold in the presence of disordered parameters $J$ and $g$ in the Hamiltonian, which enhances the prospect of experimental realization.

\subsection{Approximate calculation of the Loschmidt echo in a larger $N$}
\label{C}
In this section, we evaluate the Loschmidt echo (LE) under the post-quench
Hamiltonian $\mathcal{H}=H+\kappa D_{j}$ approximately in the two
dimensional subspace of $\left\vert \psi _{n}^{+}\right\rangle $ and $%
\left\vert \psi _{n}^{-}\right\rangle $ with the initial state $\rho
_{n}\left( 0\right) =e^{-\beta E_{n}^{+}}\left\vert \psi
_{n}^{+}\right\rangle \left\langle \psi _{n}^{+}\right\vert +e^{-\beta
	E_{n}^{-}}\left\vert \psi _{n}^{-}\right\rangle \left\langle \psi
_{n}^{-}\right\vert $. Instead of the exact calculation of the full Hilbert
space in the Letter, this allows us to see the finite-size scaling behavior
in a larger $N$. Here $\left\vert \psi _{n}^{+}\right\rangle $ and $%
\left\vert \psi _{n}^{-}\right\rangle $ are the eigenstates discussed in Eq.
(5) in the Letter, wherein they are degenerate when $g<1$ in thermodynamic
limit. Now we are considering an arbitrary $g$ in finite $N$, and the
energies for these two eigenstates are different.

In the fermion representation, the operator $D_{j}$ can be expressed as the
linear combination of $d_{n}$ and $d_{n}^{\dag }$, that is 
\begin{eqnarray}
	D_{j} &=&\left( c_{j}^{\dag }+c_{j}\right) +\left( c_{N-j+1}^{\dag
	}-c_{N-j+1}\right)  \notag \\
	&=&\sum_{n=1}^{N}\left[ A_{n}\left( j\right) d_{n}+B_{n}\left( j\right)
	d_{n}^{\dag }\right] ,
\end{eqnarray}%
where $D_{j}$, $c_{j}$ and $d_{n}$ are the operators defined in Sec. \ref{A}%
, and the $j$-dependent coefficients $A_{n}\left( j\right) $, $B_{n}\left(
j\right) $ can be obtained numerically. Here we evaluate the time evolution
operator $\exp \left( -i\mathcal{H}t\right) $. Using the Zassenhaus formula 
\cite{kimura2017explicit}, we have 
\begin{eqnarray}
	\exp \left( -i\mathcal{H}t\right) &=&\exp \left( -i\left( H+\kappa
	D_{j}\right) t\right)  \notag \\
	&=&\exp \left( -iHt\right) \exp \left( -i\kappa D_{j}t\right) \exp \left(
	\kappa \frac{t^{2}}{2}\left[ H,D_{j}\right] \right) \exp \left( i\frac{t^{3}%
	}{6}\left( 2\left[ \kappa D_{j},\left[ H,\kappa D_{j}\right] \right] +\left[
	H,\left[ H,\kappa D_{j}\right] \right] \right) \right)  \notag \\
	&&\times \exp \left( -\frac{t^{4}}{24}\left( \left[ \left[ \left[ H,\kappa
	D_{j}\right] ,H\right] ,H\right] +3\left[ \left[ \left[ H,\kappa D_{j}\right]
	,H\right] ,\kappa D_{j}\right] +3\left[ \left[ \left[ H,\kappa D_{j}\right]
	,\kappa D_{j}\right] ,\kappa D_{j}\right] \right) \right) \times ...\text{,}
\end{eqnarray}%
which can be simplified through the following process: using Eq. \ref%
{dHcommu}. and expand the terms after $\exp \left( -iHt\right) $ in Taylor
series; The terms with $d_{n\neq N}$ and $d_{n\neq N}^{\dag }$ have no
contribution to the LE in the subspace we considering, thus they can be
ignored. Finally, we obtain the time evolution operator approximately 
\begin{equation}
	\exp \left( -i\mathcal{H}t\right) \approx \exp \left( -iHt\right) \left\{
	1-\kappa \frac{A_{N}\left( j\right) }{-\varepsilon _{N}}\left[ \exp \left(
	-i\varepsilon _{N}t\right) -1\right] d_{N}-\kappa \frac{B_{N}\left( j\right) 
	}{\varepsilon _{N}}\left[ \exp \left( i\varepsilon _{N}t\right) -1\right]
	d_{N}^{\dag }\right\} .
\end{equation}

\begin{figure*}[t]
	\centering
	\includegraphics[width=1\textwidth]{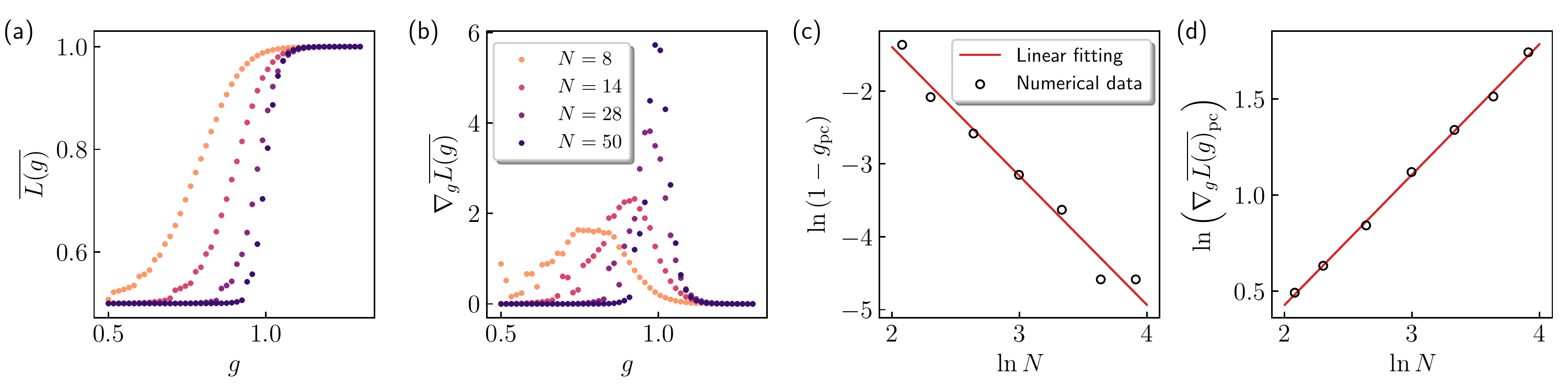}
	\caption{The numerical results of the average LEs and the finite-size
		scaling analysis when $j=1$. (a) and (b) are the average LEs and the derivatives as
		functions of $g$ for $N=8$, $14$, $28$ and $50$. (c) and (d) are $\left( g_{%
			\mathrm{c}}-g_{\mathrm{pc}}\right) $ and $\left( \protect\nabla _{g}%
		\overline{L\left( g\right) }\right) _{g=g_{\mathrm{pc}}}$ as a function of $%
		N $ in logarithmic scales, where system sizes $N=8$, $10$, $14$, $20$, $28$, 
		$38$, $50$ are taken in the numerical calculation, and the numerical data
		are fitted linearly by $\ln\left(1-g_{\mathrm{pc}}\right)=-1.77\ln N+2.14$
		and $\ln \left( \protect\nabla _g\overline{L(g)}_{\mathrm{pc}}
		\right)=0.68\ln N-0.93$. Other parameters for the numerical calculations are 
		$\protect\tau=1000$, $T=2000$, $J=1$, $\protect\kappa=0.1$, and $\protect%
		\beta=10$. }
	\label{figS1}
\end{figure*}

Having this result, it is straight forward to calculate the time evolution
of the initial state $\rho _{n}\left( 0\right) ,$ by using $d_{N}\left\vert
\psi _{n}^{+}\right\rangle =\left\vert \psi _{n}^{-}\right\rangle
,d_{N}^{\dag }\left\vert \psi _{n}^{-}\right\rangle =\left\vert \psi
_{n}^{+}\right\rangle ,d_{N}\left\vert \psi _{n}^{-}\right\rangle
=d_{N}^{\dag }\left\vert \psi _{n}^{+}\right\rangle =0$, and $\exp \left(
-iHt\right) \left\vert \psi _{n}^{\pm }\right\rangle =\exp \left(
-iE_{n}^{\pm }t\right) \left\vert \psi _{n}^{\pm }\right\rangle $. Here we
calculate the LE in the subspace of the ground state and the first-excited
state, with the initial state $\rho _{\mathrm{g}}\left( 0\right) =e^{-\beta
	E_{\mathrm{g}}^{+}}\left\vert \psi _{\mathrm{g}}^{+}\right\rangle
\left\langle \psi _{\mathrm{g}}^{+}\right\vert +e^{-\beta E_{\mathrm{g}%
	}^{-}}\left\vert \psi _{\mathrm{g}}^{-}\right\rangle \left\langle \psi _{%
	\mathrm{g}}^{-}\right\vert $. The LE in the subspace of the higher-excited
states can be calculated similarly. The numerical calculations of the LE and
the average LE follow the definitions in Eqs. (12) and (13), respectively,
in the Letter. The numerical results of the average LEs under the post-quench
Hamiltonian $\mathcal{H}=H+\kappa D_{1}$ of different system sizes are
presented in Fig. \ref{figS1}.

In Fig. \ref{figS1}(a), we plot the average LEs as functions of parameter $g$
for different $N$. Correspondingly, the derivative of the average LEs with
respect to $g$ are plotted in Fig. \ref{figS1}(b) where we can find the
pseudo critical point $g_{\mathrm{pc}}$, defined as the maximum point of $%
\nabla _{g}\overline{L\left( g\right) }$. We can see that the pseudo
critical point is closer to the critical point $g_{\mathrm{c}}=1$ for a
larger $N$. Figs. \ref{figS1}(c) and (d) are $\left( g_{\mathrm{c}}-g_{%
	\mathrm{pc}}\right) $ and $\left( \nabla _{g}\overline{L\left( g\right) }%
\right) _{g=g_{\mathrm{pc}}}$ as a function of $N$ in logarithmic scales. We
can see that the scaling behaviors are consistent to our expectation: when $%
N$ becomes larger, the pseudo critical point approaches to $1$, and the
derivative of the average LEs at the pseudo critical point tends to infinite.

The numerical results of the $j$-dependent average LEs of initial state $\rho _{\mathrm{g}}\left( 0\right) $ under the post-quench
Hamiltonian $\mathcal{H}=H+\kappa D_{j}$ are presented in Fig. \ref{figS2}. It indicates that when $g<1$, the average LEs decay with exponential law $\overline{L_{j}}=C_{1}\exp\left(C_{2}j\right)+0.5 $ close to the boundary (small $j$), where $C_{1}$ and $C_{2}$ are $g$-dependent real numbers. This suggests the bulk-boundary correspondence at nonzero temperatures in a larger $N$.

\begin{figure*}[t]
	\centering
	\includegraphics[width=1\textwidth]{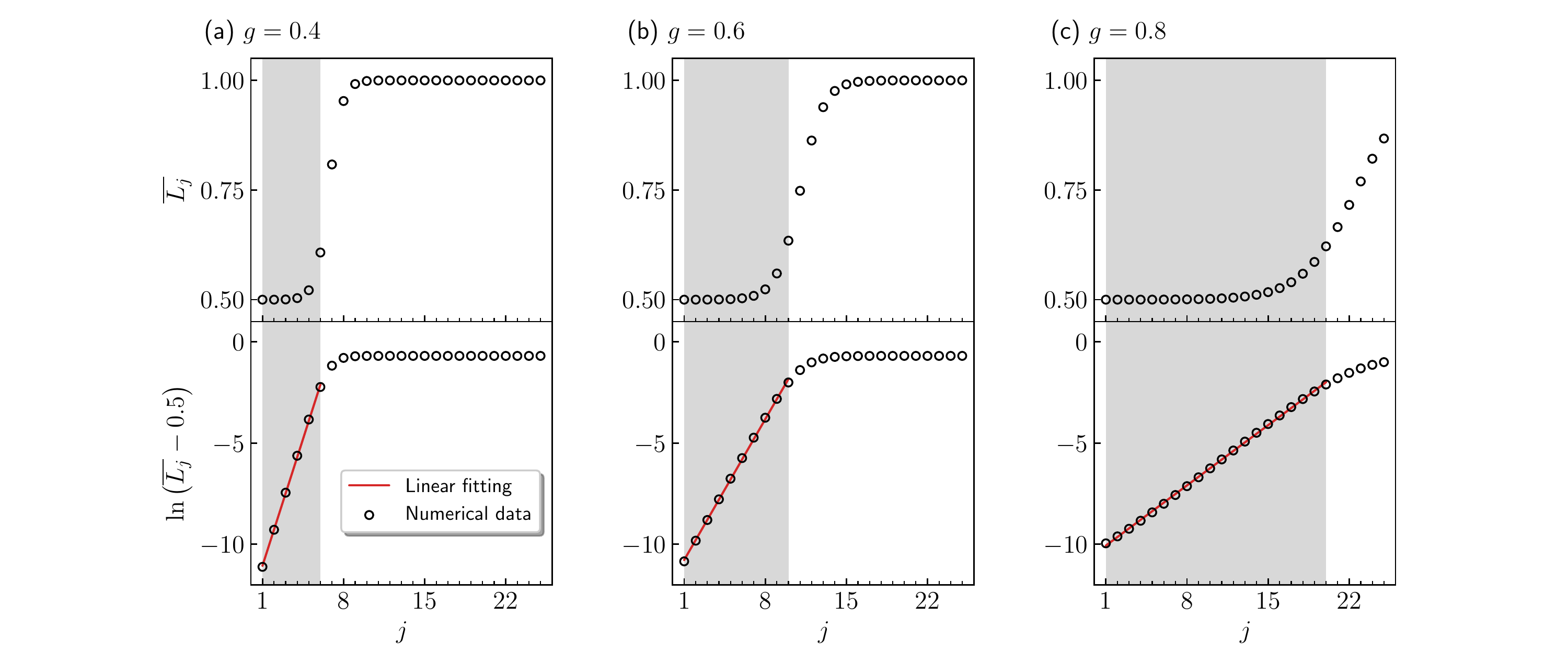}  
	\caption{The numerical results of the average LEs $\overline{L_{j}}$ as functions of $j$, for different $g$: (a) $g=0.4$; (b) $g=0.6$ and (c) $g=0.8$. In the bottom panels, $\ln \left(\overline{L_{j}}-0.5\right) $ as functions of $j$ are plotted corresponding to the upper panels. The red lines are the linear fittings for the data points in the shaded regions. Other parameters for the numerical calculations are $N=50$, 
		$\protect\tau=1000$, $T=2000$, $J=1$, $\protect\kappa=0.1$, and $\protect%
		\beta=10$.}
	\label{figS2}
\end{figure*}
\end{widetext}


\begin{thebibliography}{44}%
	\makeatletter
	\providecommand \@ifxundefined [1]{%
		\@ifx{#1\undefined}
	}%
	\providecommand \@ifnum [1]{%
		\ifnum #1\expandafter \@firstoftwo
		\else \expandafter \@secondoftwo
		\fi
	}%
	\providecommand \@ifx [1]{%
		\ifx #1\expandafter \@firstoftwo
		\else \expandafter \@secondoftwo
		\fi
	}%
	\providecommand \natexlab [1]{#1}%
	\providecommand \enquote  [1]{``#1''}%
	\providecommand \bibnamefont  [1]{#1}%
	\providecommand \bibfnamefont [1]{#1}%
	\providecommand \citenamefont [1]{#1}%
	\providecommand \href@noop [0]{\@secondoftwo}%
	\providecommand \href [0]{\begingroup \@sanitize@url \@href}%
	\providecommand \@href[1]{\@@startlink{#1}\@@href}%
	\providecommand \@@href[1]{\endgroup#1\@@endlink}%
	\providecommand \@sanitize@url [0]{\catcode `\\12\catcode `\$12\catcode
		`\&12\catcode `\#12\catcode `\^12\catcode `\_12\catcode `\%12\relax}%
	\providecommand \@@startlink[1]{}%
	\providecommand \@@endlink[0]{}%
	\providecommand \url  [0]{\begingroup\@sanitize@url \@url }%
	\providecommand \@url [1]{\endgroup\@href {#1}{\urlprefix }}%
	\providecommand \urlprefix  [0]{URL }%
	\providecommand \Eprint [0]{\href }%
	\providecommand \doibase [0]{https://doi.org/}%
	\providecommand \selectlanguage [0]{\@gobble}%
	\providecommand \bibinfo  [0]{\@secondoftwo}%
	\providecommand \bibfield  [0]{\@secondoftwo}%
	\providecommand \translation [1]{[#1]}%
	\providecommand \BibitemOpen [0]{}%
	\providecommand \bibitemStop [0]{}%
	\providecommand \bibitemNoStop [0]{.\EOS\space}%
	\providecommand \EOS [0]{\spacefactor3000\relax}%
	\providecommand \BibitemShut  [1]{\csname bibitem#1\endcsname}%
	\let\auto@bib@innerbib\@empty
	\bibitem [{\citenamefont {Sachdev}(1999)}]{sachdev1999quantum}%
	\BibitemOpen
	\bibfield  {author} {\bibinfo {author} {\bibfnamefont {S.}~\bibnamefont
			{Sachdev}},\ }\href@noop {} {\emph {\bibinfo {title} {Quantum phase
				transitions}}}\ (\bibinfo {year} {1999})\BibitemShut {NoStop}%
	\bibitem [{\citenamefont {Zhang}\ and\ \citenamefont
		{Song}(2015)}]{zhang2015topological}%
	\BibitemOpen
	\bibfield  {author} {\bibinfo {author} {\bibfnamefont {G.}~\bibnamefont
			{Zhang}}\ and\ \bibinfo {author} {\bibfnamefont {Z.}~\bibnamefont {Song}},\
	}\bibfield  {title} {\bibinfo {title} {Topological characterization of
			extended quantum ising models},\ }\href
	{https://doi.org/10.1103/PhysRevLett.115.177204} {\bibfield  {journal}
		{\bibinfo  {journal} {Phys. Rev. Lett.}\ }\textbf {\bibinfo {volume} {115}},\
		\bibinfo {pages} {177204} (\bibinfo {year} {2015})}\BibitemShut {NoStop}%
	\bibitem [{\citenamefont {Zhang}\ \emph {et~al.}(2017)\citenamefont {Zhang},
		\citenamefont {Li},\ and\ \citenamefont {Song}}]{zhang2017majorana}%
	\BibitemOpen
	\bibfield  {author} {\bibinfo {author} {\bibfnamefont {G.}~\bibnamefont
			{Zhang}}, \bibinfo {author} {\bibfnamefont {C.}~\bibnamefont {Li}},\ and\
		\bibinfo {author} {\bibfnamefont {Z.}~\bibnamefont {Song}},\ }\bibfield
	{title} {\bibinfo {title} {Majorana charges, winding numbers and chern
			numbers in quantum ising models},\ }\href
	{https://doi.org/10.1038/s41598-017-08323-0} {\bibfield  {journal} {\bibinfo
			{journal} {Sci. Rep.}\ }\textbf {\bibinfo {volume} {7}},\ \bibinfo {pages}
		{1} (\bibinfo {year} {2017})}\BibitemShut {NoStop}%
	\bibitem [{\citenamefont {Kitaev}(2001)}]{kitaev2001unpaired}%
	\BibitemOpen
	\bibfield  {author} {\bibinfo {author} {\bibfnamefont {A.~Y.}\ \bibnamefont
			{Kitaev}},\ }\bibfield  {title} {\bibinfo {title} {Unpaired majorana fermions
			in quantum wires},\ }\href {https://doi.org/10.1070/1063-7869/44/10s/s29}
	{\bibfield  {journal} {\bibinfo  {journal} {Phys. Usp.}\ }\textbf {\bibinfo
			{volume} {44}},\ \bibinfo {pages} {131} (\bibinfo {year} {2001})}\BibitemShut
	{NoStop}%
	\bibitem [{\citenamefont {Pfeuty}(1970)}]{pfeuty1970one}%
	\BibitemOpen
	\bibfield  {author} {\bibinfo {author} {\bibfnamefont {P.}~\bibnamefont
			{Pfeuty}},\ }\bibfield  {title} {\bibinfo {title} {The one-dimensional ising
			model with a transverse field},\ }\href
	{https://doi.org/https://doi.org/10.1016/0003-4916(70)90270-8} {\bibfield
		{journal} {\bibinfo  {journal} {Ann. Phys.}\ }\textbf {\bibinfo {volume}
			{57}},\ \bibinfo {pages} {79} (\bibinfo {year} {1970})}\BibitemShut {NoStop}%
	\bibitem [{\citenamefont {Su}\ \emph {et~al.}(1979)\citenamefont {Su},
		\citenamefont {Schrieffer},\ and\ \citenamefont {Heeger}}]{su1979solitons}%
	\BibitemOpen
	\bibfield  {author} {\bibinfo {author} {\bibfnamefont {W.~P.}\ \bibnamefont
			{Su}}, \bibinfo {author} {\bibfnamefont {J.~R.}\ \bibnamefont {Schrieffer}},\
		and\ \bibinfo {author} {\bibfnamefont {A.~J.}\ \bibnamefont {Heeger}},\
	}\bibfield  {title} {\bibinfo {title} {Solitons in polyacetylene},\ }\href
	{https://doi.org/10.1103/PhysRevLett.42.1698} {\bibfield  {journal} {\bibinfo
			{journal} {Phys. Rev. Lett.}\ }\textbf {\bibinfo {volume} {42}},\ \bibinfo
		{pages} {1698} (\bibinfo {year} {1979})}\BibitemShut {NoStop}%
	\bibitem [{\citenamefont {Zak}(1989)}]{zak1989berry}%
	\BibitemOpen
	\bibfield  {author} {\bibinfo {author} {\bibfnamefont {J.}~\bibnamefont
			{Zak}},\ }\bibfield  {title} {\bibinfo {title} {Berry’s phase for energy
			bands in solids},\ }\href {https://doi.org/10.1103/PhysRevLett.62.2747}
	{\bibfield  {journal} {\bibinfo  {journal} {Phys. Rev. Lett.}\ }\textbf
		{\bibinfo {volume} {62}},\ \bibinfo {pages} {2747} (\bibinfo {year}
		{1989})}\BibitemShut {NoStop}%
	\bibitem [{\citenamefont {Asb{\'o}th}\ \emph {et~al.}(2016)\citenamefont
		{Asb{\'o}th}, \citenamefont {Oroszl{\'a}ny},\ and\ \citenamefont
		{P{\'a}lyi}}]{asboth2016short}%
	\BibitemOpen
	\bibfield  {author} {\bibinfo {author} {\bibfnamefont {J.~K.}\ \bibnamefont
			{Asb{\'o}th}}, \bibinfo {author} {\bibfnamefont {L.}~\bibnamefont
			{Oroszl{\'a}ny}},\ and\ \bibinfo {author} {\bibfnamefont {A.}~\bibnamefont
			{P{\'a}lyi}},\ }\href@noop {} {\emph {\bibinfo {title} {A short course on
				topological insulators}}}\ (\bibinfo  {publisher} {Springer},\ \bibinfo
	{year} {2016})\BibitemShut {NoStop}%
	\bibitem [{\citenamefont {Dugave}\ \emph {et~al.}(2013)\citenamefont {Dugave},
		\citenamefont {G{\"o}hmann},\ and\ \citenamefont
		{Kozlowski}}]{dugave2013thermal}%
	\BibitemOpen
	\bibfield  {author} {\bibinfo {author} {\bibfnamefont {M.}~\bibnamefont
			{Dugave}}, \bibinfo {author} {\bibfnamefont {F.}~\bibnamefont
			{G{\"o}hmann}},\ and\ \bibinfo {author} {\bibfnamefont {K.~K.}\ \bibnamefont
			{Kozlowski}},\ }\bibfield  {title} {\bibinfo {title} {Thermal form factors of
			the xxz chain and the large-distance asymptotics of its temperature dependent
			correlation functions},\ }\href
	{https://doi.org/10.1088/1742-5468/2013/07/p07010} {\bibfield  {journal}
		{\bibinfo  {journal} {J. Stat. Mech.}\ }\textbf {\bibinfo {volume} {2013}},\
		\bibinfo {pages} {P07010} (\bibinfo {year} {2013})}\BibitemShut {NoStop}%
	\bibitem [{\citenamefont {Granet}\ \emph {et~al.}(2020)\citenamefont {Granet},
		\citenamefont {Fagotti},\ and\ \citenamefont {Essler}}]{granet2020finite}%
	\BibitemOpen
	\bibfield  {author} {\bibinfo {author} {\bibfnamefont {E.}~\bibnamefont
			{Granet}}, \bibinfo {author} {\bibfnamefont {M.}~\bibnamefont {Fagotti}},\
		and\ \bibinfo {author} {\bibfnamefont {F.~H.~L.}\ \bibnamefont {Essler}},\
	}\bibfield  {title} {\bibinfo {title} {{Finite temperature and quench
				dynamics in the Transverse Field Ising Model from form factor expansions}},\
	}\href {https://doi.org/10.21468/SciPostPhys.9.3.033} {\bibfield  {journal}
		{\bibinfo  {journal} {SciPost Phys.}\ }\textbf {\bibinfo {volume} {9}},\
		\bibinfo {pages} {33} (\bibinfo {year} {2020})}\BibitemShut {NoStop}%
	\bibitem [{\citenamefont {Suzuki}(1985)}]{suzuki1985transfer}%
	\BibitemOpen
	\bibfield  {author} {\bibinfo {author} {\bibfnamefont {M.}~\bibnamefont
			{Suzuki}},\ }\bibfield  {title} {\bibinfo {title} {Transfer-matrix method and
			monte carlo simulation in quantum spin systems},\ }\href
	{https://doi.org/10.1103/PhysRevB.31.2957} {\bibfield  {journal} {\bibinfo
			{journal} {Phys. Rev. B}\ }\textbf {\bibinfo {volume} {31}},\ \bibinfo
		{pages} {2957} (\bibinfo {year} {1985})}\BibitemShut {NoStop}%
	\bibitem [{\citenamefont {Andraschko}\ and\ \citenamefont
		{Sirker}(2014)}]{andraschko2014dynamical}%
	\BibitemOpen
	\bibfield  {author} {\bibinfo {author} {\bibfnamefont {F.}~\bibnamefont
			{Andraschko}}\ and\ \bibinfo {author} {\bibfnamefont {J.}~\bibnamefont
			{Sirker}},\ }\bibfield  {title} {\bibinfo {title} {Dynamical quantum phase
			transitions and the loschmidt echo: A transfer matrix approach},\ }\href
	{https://doi.org/10.1103/PhysRevB.89.125120} {\bibfield  {journal} {\bibinfo
			{journal} {Phys. Rev. B}\ }\textbf {\bibinfo {volume} {89}},\ \bibinfo
		{pages} {125120} (\bibinfo {year} {2014})}\BibitemShut {NoStop}%
	\bibitem [{\citenamefont {Quan}\ \emph {et~al.}(2006)\citenamefont {Quan},
		\citenamefont {Song}, \citenamefont {Liu}, \citenamefont {Zanardi},\ and\
		\citenamefont {Sun}}]{quan2006decay}%
	\BibitemOpen
	\bibfield  {author} {\bibinfo {author} {\bibfnamefont {H.~T.}\ \bibnamefont
			{Quan}}, \bibinfo {author} {\bibfnamefont {Z.}~\bibnamefont {Song}}, \bibinfo
		{author} {\bibfnamefont {X.~F.}\ \bibnamefont {Liu}}, \bibinfo {author}
		{\bibfnamefont {P.}~\bibnamefont {Zanardi}},\ and\ \bibinfo {author}
		{\bibfnamefont {C.~P.}\ \bibnamefont {Sun}},\ }\bibfield  {title} {\bibinfo
		{title} {Decay of loschmidt echo enhanced by quantum criticality},\ }\href
	{https://doi.org/10.1103/PhysRevLett.96.140604} {\bibfield  {journal}
		{\bibinfo  {journal} {Phys. Rev. Lett.}\ }\textbf {\bibinfo {volume} {96}},\
		\bibinfo {pages} {140604} (\bibinfo {year} {2006})}\BibitemShut {NoStop}%
	\bibitem [{\citenamefont {Zanardi}\ \emph {et~al.}(2007)\citenamefont
		{Zanardi}, \citenamefont {Quan}, \citenamefont {Wang},\ and\ \citenamefont
		{Sun}}]{zanardi2007mixed}%
	\BibitemOpen
	\bibfield  {author} {\bibinfo {author} {\bibfnamefont {P.}~\bibnamefont
			{Zanardi}}, \bibinfo {author} {\bibfnamefont {H.~T.}\ \bibnamefont {Quan}},
		\bibinfo {author} {\bibfnamefont {X.}~\bibnamefont {Wang}},\ and\ \bibinfo
		{author} {\bibfnamefont {C.~P.}\ \bibnamefont {Sun}},\ }\bibfield  {title}
	{\bibinfo {title} {Mixed-state fidelity and quantum criticality at finite
			temperature},\ }\href {https://doi.org/10.1103/PhysRevA.75.032109} {\bibfield
		{journal} {\bibinfo  {journal} {Phys. Rev. A}\ }\textbf {\bibinfo {volume}
			{75}},\ \bibinfo {pages} {032109} (\bibinfo {year} {2007})}\BibitemShut
	{NoStop}%
	\bibitem [{\citenamefont {Cozzini}\ \emph {et~al.}(2007)\citenamefont
		{Cozzini}, \citenamefont {Giorda},\ and\ \citenamefont
		{Zanardi}}]{cozzini2007quantum}%
	\BibitemOpen
	\bibfield  {author} {\bibinfo {author} {\bibfnamefont {M.}~\bibnamefont
			{Cozzini}}, \bibinfo {author} {\bibfnamefont {P.}~\bibnamefont {Giorda}},\
		and\ \bibinfo {author} {\bibfnamefont {P.}~\bibnamefont {Zanardi}},\
	}\bibfield  {title} {\bibinfo {title} {Quantum phase transitions and quantum
			fidelity in free fermion graphs},\ }\href
	{https://doi.org/10.1103/PhysRevB.75.014439} {\bibfield  {journal} {\bibinfo
			{journal} {Phys. Rev. B}\ }\textbf {\bibinfo {volume} {75}},\ \bibinfo
		{pages} {014439} (\bibinfo {year} {2007})}\BibitemShut {NoStop}%
	\bibitem [{\citenamefont {Heyl}\ \emph {et~al.}(2013)\citenamefont {Heyl},
		\citenamefont {Polkovnikov},\ and\ \citenamefont
		{Kehrein}}]{heyl2013dynamical}%
	\BibitemOpen
	\bibfield  {author} {\bibinfo {author} {\bibfnamefont {M.}~\bibnamefont
			{Heyl}}, \bibinfo {author} {\bibfnamefont {A.}~\bibnamefont {Polkovnikov}},\
		and\ \bibinfo {author} {\bibfnamefont {S.}~\bibnamefont {Kehrein}},\
	}\bibfield  {title} {\bibinfo {title} {Dynamical quantum phase transitions in
			the transverse-field ising model},\ }\href
	{https://doi.org/10.1103/PhysRevLett.110.135704} {\bibfield  {journal}
		{\bibinfo  {journal} {Phys. Rev. Lett.}\ }\textbf {\bibinfo {volume} {110}},\
		\bibinfo {pages} {135704} (\bibinfo {year} {2013})}\BibitemShut {NoStop}%
	\bibitem [{\citenamefont {Abeling}\ and\ \citenamefont
		{Kehrein}(2016)}]{abeling2016quantum}%
	\BibitemOpen
	\bibfield  {author} {\bibinfo {author} {\bibfnamefont {N.~O.}\ \bibnamefont
			{Abeling}}\ and\ \bibinfo {author} {\bibfnamefont {S.}~\bibnamefont
			{Kehrein}},\ }\bibfield  {title} {\bibinfo {title} {Quantum quench dynamics
			in the transverse field ising model at nonzero temperatures},\ }\href
	{https://doi.org/10.1103/PhysRevB.93.104302} {\bibfield  {journal} {\bibinfo
			{journal} {Phys. Rev. B}\ }\textbf {\bibinfo {volume} {93}},\ \bibinfo
		{pages} {104302} (\bibinfo {year} {2016})}\BibitemShut {NoStop}%
	\bibitem [{\citenamefont {Jafari}\ and\ \citenamefont
		{Johannesson}(2017)}]{jafari2017loschmidt}%
	\BibitemOpen
	\bibfield  {author} {\bibinfo {author} {\bibfnamefont {R.}~\bibnamefont
			{Jafari}}\ and\ \bibinfo {author} {\bibfnamefont {H.}~\bibnamefont
			{Johannesson}},\ }\bibfield  {title} {\bibinfo {title} {Loschmidt echo
			revivals: Critical and noncritical},\ }\href
	{https://doi.org/10.1103/PhysRevLett.118.015701} {\bibfield  {journal}
		{\bibinfo  {journal} {Phys. Rev. Lett.}\ }\textbf {\bibinfo {volume} {118}},\
		\bibinfo {pages} {015701} (\bibinfo {year} {2017})}\BibitemShut {NoStop}%
	\bibitem [{\citenamefont {Mera}\ \emph {et~al.}(2018)\citenamefont {Mera},
		\citenamefont {Vlachou}, \citenamefont {Paunkovi\ifmmode~\acute{c}\else
			\'{c}\fi{}}, \citenamefont {Vieira},\ and\ \citenamefont
		{Viyuela}}]{mera2018dynamical}%
	\BibitemOpen
	\bibfield  {author} {\bibinfo {author} {\bibfnamefont {B.}~\bibnamefont
			{Mera}}, \bibinfo {author} {\bibfnamefont {C.}~\bibnamefont {Vlachou}},
		\bibinfo {author} {\bibfnamefont {N.}~\bibnamefont
			{Paunkovi\ifmmode~\acute{c}\else \'{c}\fi{}}}, \bibinfo {author}
		{\bibfnamefont {V.~R.}\ \bibnamefont {Vieira}},\ and\ \bibinfo {author}
		{\bibfnamefont {O.}~\bibnamefont {Viyuela}},\ }\bibfield  {title} {\bibinfo
		{title} {Dynamical phase transitions at finite temperature from fidelity and
			interferometric loschmidt echo induced metrics},\ }\href
	{https://doi.org/10.1103/PhysRevB.97.094110} {\bibfield  {journal} {\bibinfo
			{journal} {Phys. Rev. B}\ }\textbf {\bibinfo {volume} {97}},\ \bibinfo
		{pages} {094110} (\bibinfo {year} {2018})}\BibitemShut {NoStop}%
	\bibitem [{\citenamefont {Mostafazadeh}(2002)}]{mostafazadeh2002pseudo}%
	\BibitemOpen
	\bibfield  {author} {\bibinfo {author} {\bibfnamefont {A.}~\bibnamefont
			{Mostafazadeh}},\ }\bibfield  {title} {\bibinfo {title} {Pseudo-hermiticity
			versus pt symmetry: the necessary condition for the reality of the spectrum
			of a non-hermitian hamiltonian},\ }\href {https://doi.org/10.1063/1.1418246}
	{\bibfield  {journal} {\bibinfo  {journal} {J. Math. Phys.}\ }\textbf
		{\bibinfo {volume} {43}},\ \bibinfo {pages} {205} (\bibinfo {year}
		{2002})}\BibitemShut {NoStop}%
	\bibitem [{\citenamefont {Bender}\ \emph {et~al.}(2002)\citenamefont {Bender},
		\citenamefont {Brody},\ and\ \citenamefont {Jones}}]{bender2002complex}%
	\BibitemOpen
	\bibfield  {author} {\bibinfo {author} {\bibfnamefont {C.~M.}\ \bibnamefont
			{Bender}}, \bibinfo {author} {\bibfnamefont {D.~C.}\ \bibnamefont {Brody}},\
		and\ \bibinfo {author} {\bibfnamefont {H.~F.}\ \bibnamefont {Jones}},\
	}\bibfield  {title} {\bibinfo {title} {Complex extension of quantum
			mechanics},\ }\href {https://doi.org/10.1103/PhysRevLett.89.270401}
	{\bibfield  {journal} {\bibinfo  {journal} {Phys. Rev. Lett.}\ }\textbf
		{\bibinfo {volume} {89}},\ \bibinfo {pages} {270401} (\bibinfo {year}
		{2002})}\BibitemShut {NoStop}%
	\bibitem [{\citenamefont {Bender}\ and\ \citenamefont
		{Boettcher}(1998)}]{bender1998real}%
	\BibitemOpen
	\bibfield  {author} {\bibinfo {author} {\bibfnamefont {C.~M.}\ \bibnamefont
			{Bender}}\ and\ \bibinfo {author} {\bibfnamefont {S.}~\bibnamefont
			{Boettcher}},\ }\bibfield  {title} {\bibinfo {title} {Real spectra in
			non-hermitian hamiltonians having pt symmetry},\ }\href
	{https://doi.org/10.1103/PhysRevLett.80.5243} {\bibfield  {journal} {\bibinfo
			{journal} {Phys. Rev. Lett.}\ }\textbf {\bibinfo {volume} {80}},\ \bibinfo
		{pages} {5243} (\bibinfo {year} {1998})}\BibitemShut {NoStop}%
	\bibitem [{\citenamefont {Bender}\ \emph {et~al.}(1999)\citenamefont {Bender},
		\citenamefont {Boettcher},\ and\ \citenamefont {Meisinger}}]{bender1999pt}%
	\BibitemOpen
	\bibfield  {author} {\bibinfo {author} {\bibfnamefont {C.~M.}\ \bibnamefont
			{Bender}}, \bibinfo {author} {\bibfnamefont {S.}~\bibnamefont {Boettcher}},\
		and\ \bibinfo {author} {\bibfnamefont {P.~N.}\ \bibnamefont {Meisinger}},\
	}\bibfield  {title} {\bibinfo {title} {Pt-symmetric quantum mechanics},\
	}\href {https://doi.org/10.1063/1.532860} {\bibfield  {journal} {\bibinfo
			{journal} {J. Math. Phys.}\ }\textbf {\bibinfo {volume} {40}},\ \bibinfo
		{pages} {2201} (\bibinfo {year} {1999})}\BibitemShut {NoStop}%
	\bibitem [{\citenamefont {Mostafazadeh}(2009)}]{mostafazadeh2009spectral}%
	\BibitemOpen
	\bibfield  {author} {\bibinfo {author} {\bibfnamefont {A.}~\bibnamefont
			{Mostafazadeh}},\ }\bibfield  {title} {\bibinfo {title} {Spectral
			singularities of complex scattering potentials and infinite reflection and
			transmission coefficients at real energies},\ }\href
	{https://doi.org/10.1103/PhysRevLett.102.220402} {\bibfield  {journal}
		{\bibinfo  {journal} {Phys. Rev. Lett.}\ }\textbf {\bibinfo {volume} {102}},\
		\bibinfo {pages} {220402} (\bibinfo {year} {2009})}\BibitemShut {NoStop}%
	\bibitem [{\citenamefont {Longhi}(2014)}]{longhi2014exceptional}%
	\BibitemOpen
	\bibfield  {author} {\bibinfo {author} {\bibfnamefont {S.}~\bibnamefont
			{Longhi}},\ }\bibfield  {title} {\bibinfo {title} {Exceptional points and
			bloch oscillations in non-hermitian lattices with unidirectional hopping},\
	}\href {https://doi.org/10.1209/0295-5075/106/34001} {\bibfield  {journal}
		{\bibinfo  {journal} {{EPL} (Europhysics Letters)}\ }\textbf {\bibinfo
			{volume} {106}},\ \bibinfo {pages} {34001} (\bibinfo {year}
		{2014})}\BibitemShut {NoStop}%
	\bibitem [{\citenamefont {Jin}\ and\ \citenamefont
		{Song}(2018)}]{jin2018incident}%
	\BibitemOpen
	\bibfield  {author} {\bibinfo {author} {\bibfnamefont {L.}~\bibnamefont
			{Jin}}\ and\ \bibinfo {author} {\bibfnamefont {Z.}~\bibnamefont {Song}},\
	}\bibfield  {title} {\bibinfo {title} {Incident direction independent wave
			propagation and unidirectional lasing},\ }\href
	{https://doi.org/10.1103/PhysRevLett.121.073901} {\bibfield  {journal}
		{\bibinfo  {journal} {Phys. Rev. Lett.}\ }\textbf {\bibinfo {volume} {121}},\
		\bibinfo {pages} {073901} (\bibinfo {year} {2018})}\BibitemShut {NoStop}%
	\bibitem [{\citenamefont {Zhang}\ \emph {et~al.}(2020)\citenamefont {Zhang},
		\citenamefont {Jin},\ and\ \citenamefont {Song}}]{zhang2020dynamic}%
	\BibitemOpen
	\bibfield  {author} {\bibinfo {author} {\bibfnamefont {X.~Z.}\ \bibnamefont
			{Zhang}}, \bibinfo {author} {\bibfnamefont {L.}~\bibnamefont {Jin}},\ and\
		\bibinfo {author} {\bibfnamefont {Z.}~\bibnamefont {Song}},\ }\bibfield
	{title} {\bibinfo {title} {Dynamic magnetization in non-hermitian quantum
			spin systems},\ }\href {https://doi.org/10.1103/PhysRevB.101.224301}
	{\bibfield  {journal} {\bibinfo  {journal} {Phys. Rev. B}\ }\textbf {\bibinfo
			{volume} {101}},\ \bibinfo {pages} {224301} (\bibinfo {year}
		{2020})}\BibitemShut {NoStop}%
	\bibitem [{\citenamefont {Dalibard}\ \emph {et~al.}(1992)\citenamefont
		{Dalibard}, \citenamefont {Castin},\ and\ \citenamefont
		{M{\o}lmer}}]{dalibard1992wave}%
	\BibitemOpen
	\bibfield  {author} {\bibinfo {author} {\bibfnamefont {J.}~\bibnamefont
			{Dalibard}}, \bibinfo {author} {\bibfnamefont {Y.}~\bibnamefont {Castin}},\
		and\ \bibinfo {author} {\bibfnamefont {K.}~\bibnamefont {M{\o}lmer}},\
	}\bibfield  {title} {\bibinfo {title} {Wave-function approach to dissipative
			processes in quantum optics},\ }\href
	{https://doi.org/10.1103/PhysRevLett.68.580} {\bibfield  {journal} {\bibinfo
			{journal} {Phys. Rev. Lett.}\ }\textbf {\bibinfo {volume} {68}},\ \bibinfo
		{pages} {580} (\bibinfo {year} {1992})}\BibitemShut {NoStop}%
	\bibitem [{\citenamefont {Dum}\ \emph {et~al.}(1992)\citenamefont {Dum},
		\citenamefont {Zoller},\ and\ \citenamefont {Ritsch}}]{dum1992monte}%
	\BibitemOpen
	\bibfield  {author} {\bibinfo {author} {\bibfnamefont {R.}~\bibnamefont
			{Dum}}, \bibinfo {author} {\bibfnamefont {P.}~\bibnamefont {Zoller}},\ and\
		\bibinfo {author} {\bibfnamefont {H.}~\bibnamefont {Ritsch}},\ }\bibfield
	{title} {\bibinfo {title} {Monte carlo simulation of the atomic master
			equation for spontaneous emission},\ }\href
	{https://doi.org/10.1103/PhysRevA.45.4879} {\bibfield  {journal} {\bibinfo
			{journal} {Phys. Rev. A}\ }\textbf {\bibinfo {volume} {45}},\ \bibinfo
		{pages} {4879} (\bibinfo {year} {1992})}\BibitemShut {NoStop}%
	\bibitem [{\citenamefont {M{\o}lmer}\ \emph {et~al.}(1993)\citenamefont
		{M{\o}lmer}, \citenamefont {Castin},\ and\ \citenamefont
		{Dalibard}}]{molmer1993monte}%
	\BibitemOpen
	\bibfield  {author} {\bibinfo {author} {\bibfnamefont {K.}~\bibnamefont
			{M{\o}lmer}}, \bibinfo {author} {\bibfnamefont {Y.}~\bibnamefont {Castin}},\
		and\ \bibinfo {author} {\bibfnamefont {J.}~\bibnamefont {Dalibard}},\
	}\bibfield  {title} {\bibinfo {title} {Monte carlo wave-function method in
			quantum optics},\ }\href {https://doi.org/10.1364/JOSAB.10.000524} {\bibfield
		{journal} {\bibinfo  {journal} {J. Opt. Soc. Am. B}\ }\textbf {\bibinfo
			{volume} {10}},\ \bibinfo {pages} {524} (\bibinfo {year} {1993})}\BibitemShut
	{NoStop}%
	\bibitem [{\citenamefont {Wiseman}(1996)}]{wiseman1996quantum}%
	\BibitemOpen
	\bibfield  {author} {\bibinfo {author} {\bibfnamefont {H.~M.}\ \bibnamefont
			{Wiseman}},\ }\bibfield  {title} {\bibinfo {title} {Quantum trajectories and
			quantum measurement theory},\ }\href
	{https://doi.org/10.1088/1355-5111/8/1/015} {\bibfield  {journal} {\bibinfo
			{journal} {Quantum Semiclass. Opt.}\ }\textbf {\bibinfo {volume} {8}},\
		\bibinfo {pages} {205} (\bibinfo {year} {1996})}\BibitemShut {NoStop}%
	\bibitem [{\citenamefont {Plenio}\ and\ \citenamefont
		{Knight}(1998)}]{plenio1998quantum}%
	\BibitemOpen
	\bibfield  {author} {\bibinfo {author} {\bibfnamefont {M.~B.}\ \bibnamefont
			{Plenio}}\ and\ \bibinfo {author} {\bibfnamefont {P.~L.}\ \bibnamefont
			{Knight}},\ }\bibfield  {title} {\bibinfo {title} {The quantum-jump approach
			to dissipative dynamics in quantum optics},\ }\href
	{https://doi.org/10.1103/RevModPhys.70.101} {\bibfield  {journal} {\bibinfo
			{journal} {Rev. Mod. Phys.}\ }\textbf {\bibinfo {volume} {70}},\ \bibinfo
		{pages} {101} (\bibinfo {year} {1998})}\BibitemShut {NoStop}%
	\bibitem [{\citenamefont {Lee}\ and\ \citenamefont
		{Chan}(2014)}]{lee2014heralded}%
	\BibitemOpen
	\bibfield  {author} {\bibinfo {author} {\bibfnamefont {T.~E.}\ \bibnamefont
			{Lee}}\ and\ \bibinfo {author} {\bibfnamefont {C.-K.}\ \bibnamefont {Chan}},\
	}\bibfield  {title} {\bibinfo {title} {Heralded magnetism in non-hermitian
			atomic systems},\ }\href {https://doi.org/10.1103/PhysRevX.4.041001}
	{\bibfield  {journal} {\bibinfo  {journal} {Phys. Rev. X}\ }\textbf {\bibinfo
			{volume} {4}},\ \bibinfo {pages} {041001} (\bibinfo {year}
		{2014})}\BibitemShut {NoStop}%
	\bibitem [{\citenamefont {Sachdev}\ and\ \citenamefont
		{Young}(1997)}]{sachdev1997low}%
	\BibitemOpen
	\bibfield  {author} {\bibinfo {author} {\bibfnamefont {S.}~\bibnamefont
			{Sachdev}}\ and\ \bibinfo {author} {\bibfnamefont {A.~P.}\ \bibnamefont
			{Young}},\ }\bibfield  {title} {\bibinfo {title} {Low temperature
			relaxational dynamics of the ising chain in a transverse field},\ }\href
	{https://doi.org/10.1103/PhysRevLett.78.2220} {\bibfield  {journal} {\bibinfo
			{journal} {Phys. Rev. Lett.}\ }\textbf {\bibinfo {volume} {78}},\ \bibinfo
		{pages} {2220} (\bibinfo {year} {1997})}\BibitemShut {NoStop}%
	\bibitem [{Sup()}]{Supplement}%
	\BibitemOpen
	\bibfield  {title}  {\bibinfo {title} {See Supplemental Material for the details on derivation of the
			nonlocal operator $D$ with uniform (Sec. A) as well as disordered (Sec. B) parameters $J$ and $%
			g$, and in Sec. C approximate calculation of the Loschmidt echo in a larger $N$, which includes 
			Refs. \cite{kitaev2001unpaired, asboth2016short, jordan1993paulische, kimura2017explicit}.}}\href@noop {} {\ }\BibitemShut
	{NoStop}%
	\bibitem [{\citenamefont {Jordan}\ and\ \citenamefont
		{Wigner}(1993)}]{jordan1993paulische}%
	\BibitemOpen
	\bibfield  {author} {\bibinfo {author} {\bibfnamefont {P.}~\bibnamefont
			{Jordan}}\ and\ \bibinfo {author} {\bibfnamefont {E.~P.}\ \bibnamefont
			{Wigner}},\ }\bibinfo {title} {{\"u}ber das paulische {\"a}quivalenzverbot},\
	in\ \href {https://doi.org/10.1007/978-3-662-02781-3_9} {\emph {\bibinfo
			{booktitle} {The Collected Works of Eugene Paul Wigner}}}\ (\bibinfo
	{publisher} {Springer},\ \bibinfo {year} {1993})\ pp.\ \bibinfo {pages}
	{109--129}\BibitemShut {NoStop}%
	\bibitem [{\citenamefont {Brody}\ and\ \citenamefont
		{Graefe}(2012)}]{brody2012mixed}%
	\BibitemOpen
	\bibfield  {author} {\bibinfo {author} {\bibfnamefont {D.~C.}\ \bibnamefont
			{Brody}}\ and\ \bibinfo {author} {\bibfnamefont {E.-M.}\ \bibnamefont
			{Graefe}},\ }\bibfield  {title} {\bibinfo {title} {Mixed-state evolution in
			the presence of gain and loss},\ }\href
	{https://doi.org/10.1103/PhysRevLett.109.230405} {\bibfield  {journal}
		{\bibinfo  {journal} {Phys. Rev. Lett.}\ }\textbf {\bibinfo {volume} {109}},\
		\bibinfo {pages} {230405} (\bibinfo {year} {2012})}\BibitemShut {NoStop}%
	\bibitem [{\citenamefont {Kawabata}\ \emph {et~al.}(2017)\citenamefont
		{Kawabata}, \citenamefont {Ashida},\ and\ \citenamefont
		{Ueda}}]{kawabata2017information}%
	\BibitemOpen
	\bibfield  {author} {\bibinfo {author} {\bibfnamefont {K.}~\bibnamefont
			{Kawabata}}, \bibinfo {author} {\bibfnamefont {Y.}~\bibnamefont {Ashida}},\
		and\ \bibinfo {author} {\bibfnamefont {M.}~\bibnamefont {Ueda}},\ }\bibfield
	{title} {\bibinfo {title} {Information retrieval and criticality in
			parity-time-symmetric systems},\ }\href
	{https://doi.org/10.1103/PhysRevLett.119.190401} {\bibfield  {journal}
		{\bibinfo  {journal} {Phys. Rev. Lett.}\ }\textbf {\bibinfo {volume} {119}},\
		\bibinfo {pages} {190401} (\bibinfo {year} {2017})}\BibitemShut {NoStop}%
	\bibitem [{\citenamefont {Uhlmann}(1976)}]{uhlmann1976transition}%
	\BibitemOpen
	\bibfield  {author} {\bibinfo {author} {\bibfnamefont {A.}~\bibnamefont
			{Uhlmann}},\ }\bibfield  {title} {\bibinfo {title} {The “transition
			probability” in the state space of a *-algebra},\ }\href
	{https://doi.org/https://doi.org/10.1016/0034-4877(76)90060-4} {\bibfield
		{journal} {\bibinfo  {journal} {Rep. Math. Phys.}\ }\textbf {\bibinfo
			{volume} {9}},\ \bibinfo {pages} {273 } (\bibinfo {year} {1976})}\BibitemShut
	{NoStop}%
	\bibitem [{\citenamefont {Jacobson}\ \emph {et~al.}(2011)\citenamefont
		{Jacobson}, \citenamefont {Venuti},\ and\ \citenamefont
		{Zanardi}}]{jacobson2011unitary}%
	\BibitemOpen
	\bibfield  {author} {\bibinfo {author} {\bibfnamefont {N.~T.}\ \bibnamefont
			{Jacobson}}, \bibinfo {author} {\bibfnamefont {L.~C.}\ \bibnamefont
			{Venuti}},\ and\ \bibinfo {author} {\bibfnamefont {P.}~\bibnamefont
			{Zanardi}},\ }\bibfield  {title} {\bibinfo {title} {Unitary equilibration
			after a quantum quench of a thermal state},\ }\href
	{https://doi.org/10.1103/PhysRevA.84.022115} {\bibfield  {journal} {\bibinfo
			{journal} {Phys. Rev. A}\ }\textbf {\bibinfo {volume} {84}},\ \bibinfo
		{pages} {022115} (\bibinfo {year} {2011})}\BibitemShut {NoStop}%
	\bibitem [{Sup()}]{SupplementC}%
	\BibitemOpen
	\bibfield  {title}  {\bibinfo {title} {This is also verified by the approximate calculation 
			of LEs in a larger $N$ (See Sec. C of the Supplemental Material \cite{Supplement}). 
			It indicates that when $g<1$, the average LEs decays with exponential law 
			close to the boundary (small $j$). 
			A finite-size scaling based on the approximate calculation is also given.}}\href@noop {} {\ }\BibitemShut
	{NoStop}%
	\bibitem [{\citenamefont {Huang}\ \emph {et~al.}(2011)\citenamefont {Huang},
		\citenamefont {Kong}, \citenamefont {Zhao}, \citenamefont {Shi},
		\citenamefont {Wang}, \citenamefont {Rong}, \citenamefont {Liu},\ and\
		\citenamefont {Du}}]{huang2011observation}%
	\BibitemOpen
	\bibfield  {author} {\bibinfo {author} {\bibfnamefont {P.}~\bibnamefont
			{Huang}}, \bibinfo {author} {\bibfnamefont {X.}~\bibnamefont {Kong}},
		\bibinfo {author} {\bibfnamefont {N.}~\bibnamefont {Zhao}}, \bibinfo {author}
		{\bibfnamefont {F.}~\bibnamefont {Shi}}, \bibinfo {author} {\bibfnamefont
			{P.}~\bibnamefont {Wang}}, \bibinfo {author} {\bibfnamefont {X.}~\bibnamefont
			{Rong}}, \bibinfo {author} {\bibfnamefont {R.-B.}\ \bibnamefont {Liu}},\ and\
		\bibinfo {author} {\bibfnamefont {J.}~\bibnamefont {Du}},\ }\bibfield
	{title} {\bibinfo {title} {Observation of an anomalous decoherence effect in
			a quantum bath at room temperature},\ }\href
	{https://doi.org/10.1038/ncomms1579} {\bibfield  {journal} {\bibinfo
			{journal} {Nat. Commun.}\ }\textbf {\bibinfo {volume} {2}},\ \bibinfo {pages}
		{1} (\bibinfo {year} {2011})}\BibitemShut {NoStop}%
	\bibitem [{\citenamefont {S{\'a}nchez}\ \emph {et~al.}(2020)\citenamefont
		{S{\'a}nchez}, \citenamefont {Chattah}, \citenamefont {Wei}, \citenamefont
		{Buljubasich}, \citenamefont {Cappellaro},\ and\ \citenamefont
		{Pastawski}}]{sanchez2020perturbation}%
	\BibitemOpen
	\bibfield  {author} {\bibinfo {author} {\bibfnamefont {C.~M.}\ \bibnamefont
			{S{\'a}nchez}}, \bibinfo {author} {\bibfnamefont {A.~K.}\ \bibnamefont
			{Chattah}}, \bibinfo {author} {\bibfnamefont {K.~X.}\ \bibnamefont {Wei}},
		\bibinfo {author} {\bibfnamefont {L.}~\bibnamefont {Buljubasich}}, \bibinfo
		{author} {\bibfnamefont {P.}~\bibnamefont {Cappellaro}},\ and\ \bibinfo
		{author} {\bibfnamefont {H.~M.}\ \bibnamefont {Pastawski}},\ }\bibfield
	{title} {\bibinfo {title} {Perturbation independent decay of the loschmidt
			echo in a many-body system},\ }\href
	{https://doi.org/10.1103/PhysRevLett.124.030601} {\bibfield  {journal}
		{\bibinfo  {journal} {Phys. Rev. Lett.}\ }\textbf {\bibinfo {volume} {124}},\
		\bibinfo {pages} {030601} (\bibinfo {year} {2020})}\BibitemShut {NoStop}%
	\bibitem [{\citenamefont {Kimura}(2017)}]{kimura2017explicit}%
	\BibitemOpen
	\bibfield  {author} {\bibinfo {author} {\bibfnamefont {T.}~\bibnamefont
			{Kimura}},\ }\bibfield  {title} {\bibinfo {title} {Explicit description of
			the zassenhaus formula},\ }\href
	{https://doi.org/https://doi.org/10.1093/ptep/ptx044} {\bibfield  {journal}
		{\bibinfo  {journal} {Prog. Theor. Exp. Phys.}\ }\textbf {\bibinfo {volume}
			{2017}},\ \bibinfo {pages} {041A03} (\bibinfo {year} {2017})}\BibitemShut
	{NoStop}%
\end{thebibliography}
\end{document}